\documentclass[%reprint,
superscriptaddress,
 amsmath,amssymb,
 aps,
prb,two column
]{revtex4-1}

\usepackage{graphicx}% Include figure files
\usepackage{dcolumn}% Align table columns on decimal point
\usepackage{bm}% bold math
\usepackage{graphicx,epsf,bm,bbm,color,amsmath,ulem,amssymb,float, subfigure}

\usepackage{color}
\usepackage{ textcomp }
\usepackage{ dsfont }
\usepackage{indentfirst}
\newcommand{\remove}[1]{}

\newcommand{\bi}{\begin{itemize}}
\newcommand{\ei}{\end{itemize}}
\newcommand{\be}{\begin{enumerate}}
\newcommand{\ee}{\end{enumerate}}
\newenvironment{dfn}{{\vspace*{1ex} \noindent \bf Definition }}{\vspace*{1ex}}

	\newcommand{\beq}{\begin{eqnarray}}
	\newcommand{\eeq}{\end{eqnarray}}

\begin{document}

%\preprint{APS/123-QED}

\title{Nodeless $s$-Wave Superconducting Phases in Cuprates with Ba$_2$CuO$_3$-Type Structure}

\author{Zhi-Qiang Gao}
\altaffiliation{These two authors contribute equally to this work.}
\affiliation{International Center for Quantum Materials, School of Physics, Peking University, Beijing 100871, China
}

\author{Kai-Wei Sun}
\altaffiliation{These two authors contribute equally to this work.}
\affiliation{International Center for Quantum Materials, School of Physics, Peking University, Beijing 100871, China
}

\author{Fa Wang}
\affiliation{International Center for Quantum Materials, School of Physics, Peking University, Beijing 100871, China
}
\affiliation{Collaborative Innovation Center of Quantum Matter, Beijing 100871, China
}

\date{\today}% It is always \today, today,

\begin{abstract}
In this work zero-temperature phase diagrams of cuprates with Ba$_2$CuO$_3$-type CuO chain structure is investigated. The projective symmetry group analysis is employed in the strong coupling limit, and renormalization group with bosonization analysis is employed in the weak coupling limit. We find that in both of these two limits, large areas of the phase diagrams are filled with nodeless $s$-wave superconducting phases (with weak $d$-wave components), instead of pure $d$-wave phase mostly found in cuprates. This implies that nodeless $s$-wave phase is the dominant superconducting phase in cuprates with Ba$_2$CuO$_3$-type CuO chain structure in low temperature. Other phases are also found, including $(s+d)$-wave superconducting phases and Luttinger liquid phases.
%\begin{description}
%\end{description}
\end{abstract}

\maketitle

%\tableofcontents

\section{Introduction}

\begin{figure}[htbp]
\centering
\includegraphics[width=8cm]{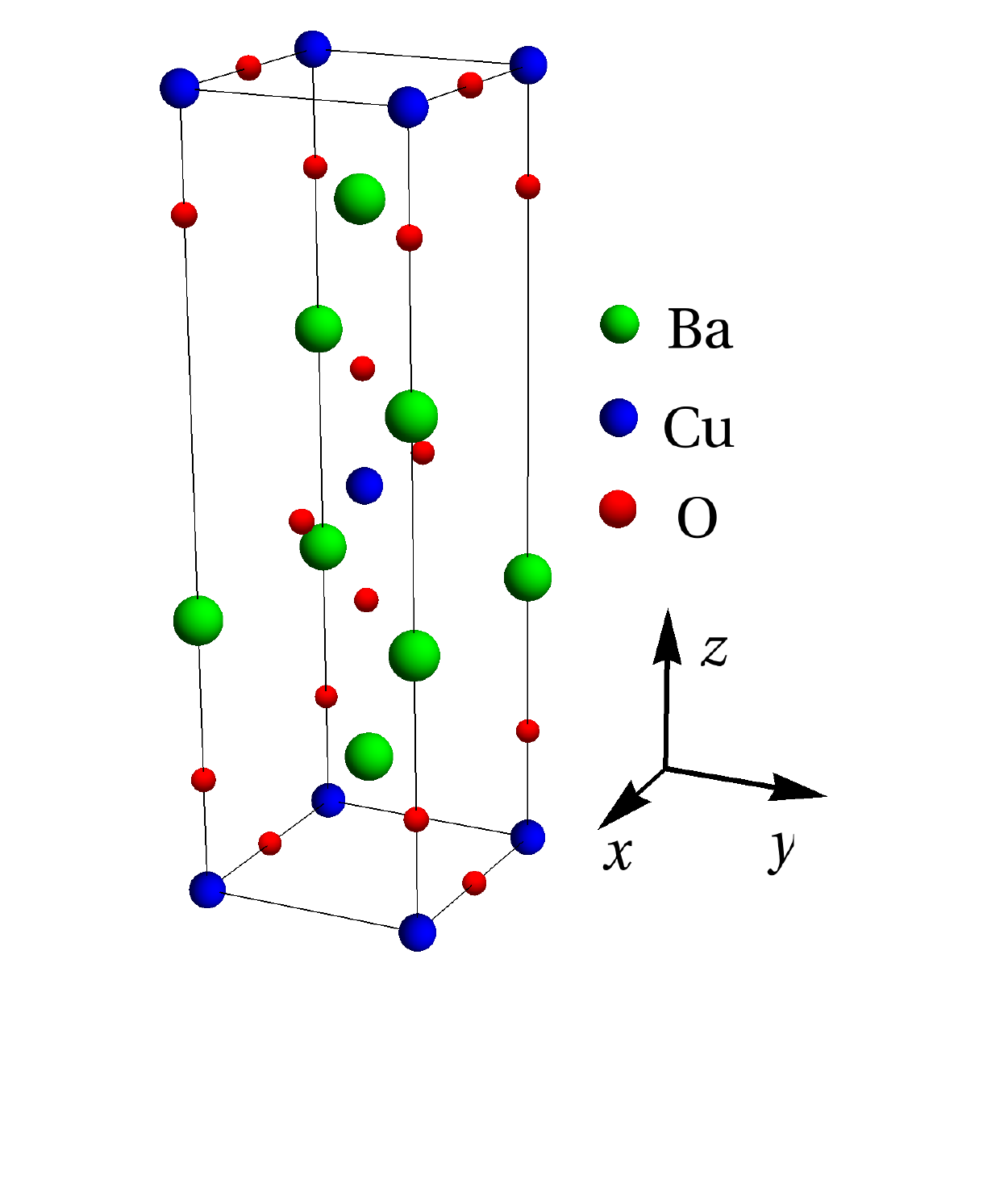}
\caption{The energetic preferred crystal structure\cite{Liu} of Ba$_2$CuO$_3$.}\label{fig:1}
\end{figure}

CuO$_2$ plane\cite{Da:1994aa,Leggett_2006} plays an important role in the cuprate superconductors with high transition temperature (high-$T_C$)\cite{Bednorz_1986,Schilling_1993,Gao:1994aa}, especially in the formation of $d$-wave pairing symmetry\cite{Ch:1993aa,Kirtley_2006,Tsuei:1994aa,Tsuei_1997}. Traditionally, oxygen vacancies in CuO$_2$ planes are detrimental\cite{Mat} to high-$T_C$. However, recent experiments\cite{Li:2019aa,Li_2019} reported that in one kind of cuprates with a large amount of oxygen vacancies, Ba$_2$CuO$_{3+\delta}$ with $0\le \delta \le 1$, high-$T_C$ superconductivity is still observed. Various theoretical works\cite{Mai:2019,Wang:2020,Li:2019aa,Le:2019,Ni:2019,Ya:2020,Zh} focusing on Ba$_2$CuO$_{3+\delta}$ have been proposed to determine the pairing symmetry and low-temperature phases in different crystal structures. Liu\cite{Liu} and coworkers showed by first principle calculation that Ba$_2$CuO$_{3+\delta}$ can be viewed as doped Ba$_2$CuO$_3$, which exhibits a CuO chain structure shown in Fig. \ref{fig:1} with one $E_g$ orbital (Cu 3$d_{z^2-x^2}$) active, and with strong intra-chain and weak inter-chain anti-ferromagnetic (AFM) coupling. In this work, we study the zero-temperature phases in Ba$_2$CuO$_{3+\delta}$ by investigating a single-orbital multi-chain t-J model in both strong and weak coupling limits.

In both limits, we find that large areas of these phase diagrams are filled with $s$-wave superconducting phases (with weak $d$-wave components). It is  $s_\pm$-wave with weak $d$-wave components (denoted as $s^d_\pm$-wave) in strong coupling limit, and $s$-wave with weak $d$-wave components (denoted as $s_d$-wave) in weak coupling limit. Both of them are nodeless on Fermi surfaces. This result indicates that the dominant superconducting phase in cuprates with Ba$_2$CuO$_3$-type CuO chain structure in low temperature is actually a nodeless $s$-wave phase, in contrast to the traditional $d$-wave phase in cuprates with CuO$_2$ plane structure. This paper is organized as following. In Sec. II  the single-orbital\cite{Liu} t-J model is introduced to describe the system. In Sec. III and IV, the strong and weak coupling limits are investigated and corresponding phase diagrams are given, respectively. We draw the conclusions in Sec. V. Details are listed in Appendix.

\section{The Model}

\begin{figure}[htbp]
\centering
\includegraphics[width=9cm]{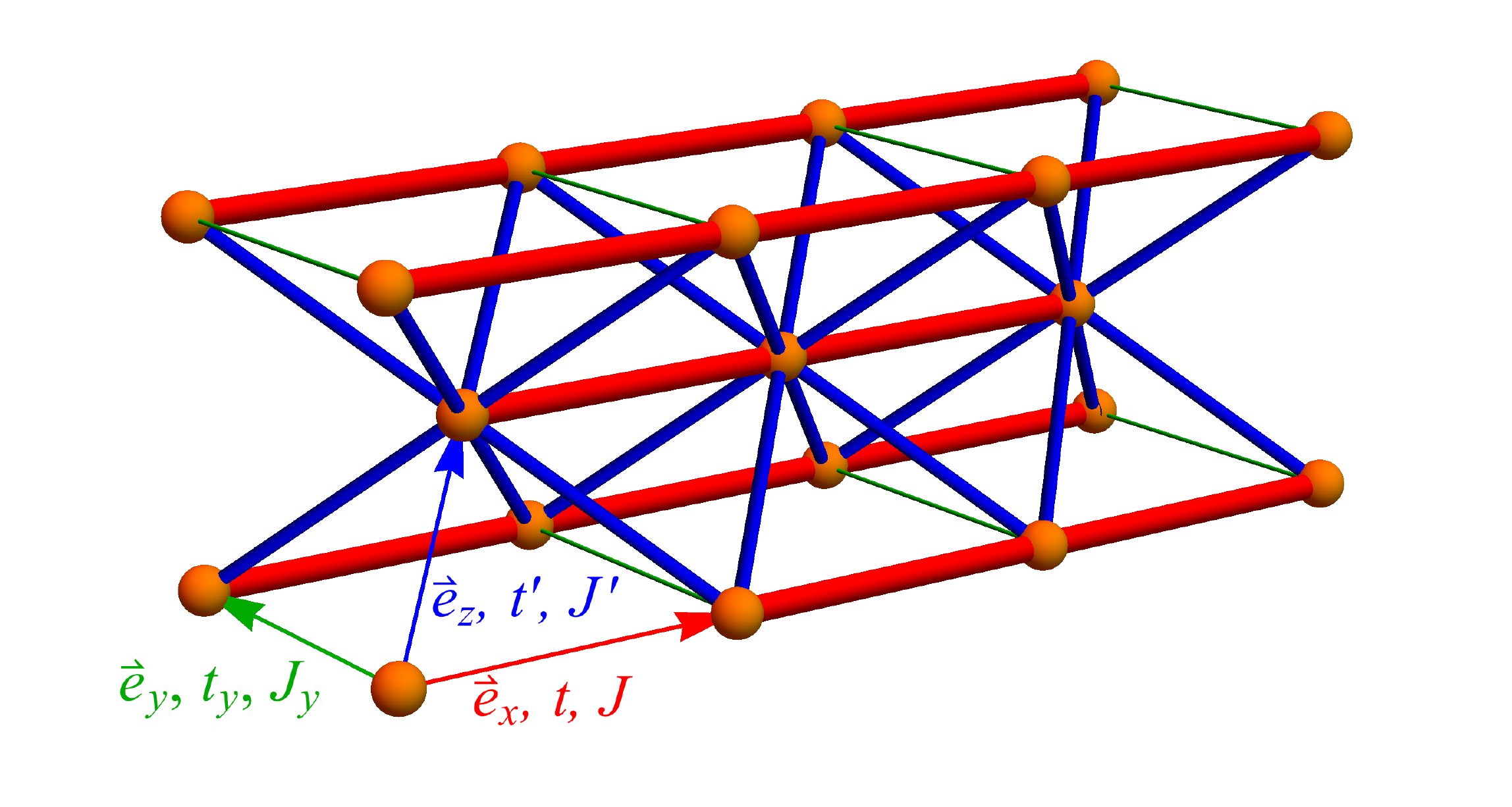}
\caption{The lattice of Ba$_2$CuO$_3$. The $x$, $y$ and $z$ direction are defined in the figure. $t,J$ are on the thickest red bonds, $t',J'$ are on the mid-thick blue bonds, and $t_y,J_y$ are on the thinnest green bonds.}\label{fig:2}
\end{figure}

As indicated in Ref.\cite{Liu}, the only active orbital is Cu 3$d_{z^2-x^2}$. Therefore, a single-orbital multi-chain t-J model is employed to describe the system. The Hamiltonian $H=H_0+H'$ reads
\beq
H_0=\sum_{x,y,z,s}&-&t(c_{x,y,z,s}^{\dag}c_{x+1,y,z,s}+h.c.)\nonumber\\
&-&t_y(c_{x,y,z,s}^{\dag}c_{x,y+1,z,s}+h.c.)\nonumber\\
&-&t'(c_{x,y,z,s}^{\dag}c_{x,y,z+1,s}+c_{x,y,z,s}^{\dag}c_{x-1,y-1,z+1,s}\nonumber\\
&+&c_{x,y,z,s}^{\dag}c_{x-1,y,z+1,s}+c_{x,y,z,s}^{\dag}c_{x,y-1,z+1,s}\nonumber\\
&+&h.c.),\label{eq:201}
\eeq
and
\beq
H'&=&\sum_{x,y,z}J\vec{S}_{x,y,z}\cdot\vec{S}_{x+1,y,z}\nonumber\\
&+&J_y\vec{S}_{x,y,z}\cdot\vec{S}_{x,y+1,z}\nonumber\\
&+&J'\vec{S}_{x,y,z}\cdot(\vec{S}_{x,y,z+1}+\vec{S}_{x-1,y,z+1}\nonumber\\
&+&\vec{S}_{x,y-1,z+1}+\vec{S}_{x-1,y-1,z+1}),\label{eq:202}
\eeq
where $c$'s ($c^\dag$'s) are annihilation (creation) operators of electrons, parameters $t,t_y,t_z,t'$ are hopping amplitudes with $t$ largest\cite{Liu}, $J,J_y,J'$ are AFM couplings with $J$ largest\cite{Liu}, and $\vec{S}_{x,y,z}=\sum_{s,s'}c_{x,y,z,s}^{\dag}\vec{\sigma}_{ss'}c_{x,y,z,s'}$ with $x$, $y$, $z$ defined in Fig. \ref{fig:2} and Pauli matrices $\vec{\sigma}=(\sigma^1,\sigma^2,\sigma^3)$. Hopping and interaction on bonds perpendicular to $xy$-plane are neglected because of the large length\cite{Liu,Li:2019aa} of these bonds.

In following sections, the single-orbital multi-chain t-J model (Eq. \ref{eq:201} and \ref{eq:202}) is studied in strong and weak coupling limits.

\section{Strong Coupling Limit}

In the strong coupling limit ($t\ll J$), projective construction (slave particle) mean field approach\cite{Wen_2002,Zhou:aa,Lee} is employed to analyze possible phases. The mean field Hamiltonian is obtained for doping description through the $SU(2)$ slave boson approach \cite{Lee}. These phases are characterized by mean field ansatzes classified by projective symmetry groups\cite{Wen_2002,Zhou:aa,Lee} (PSG's). Numerical minimization of mean field energy is employed to determine the phase diagram.

\subsection{Mean Field Hamiltonian}

A variation method is used in this section to analyze phases. A series of mean field ansatzes are introduced representing different situations of ground states and are further optimized utilizing differential evolution algorithm. 

In this approach operators of electrons are represented in spin-0 charged bosons (holons) $b_i=(b_{1,i},b_{2,i})^{\mathbf{T}}$ and spin-1/2 neutral fermions (spinons) $\psi_i=(f_{\uparrow,i},f^\dagger_{\downarrow,i})^{\mathbf{T}}$ via\cite{Lee}
\beq
c_{\uparrow,i}=\frac{1}{\sqrt{2}}b_i^\dagger \psi_i,\\
c_{\downarrow,i}=\frac{1}{\sqrt{2}}b_i^\dagger \bar{\psi}_i,
\eeq
where $\bar{\psi}_i=(f_{\downarrow,i},-f^\dagger_{\uparrow,i})^{\mathbf{T}}$. In this representation, the mean field Hamiltonian reads\cite{Lee}
\beq
H_{\text{MF}}&=&\frac{3}{4}\sum_{<ij>}J_{ij}(\text{tr}(u_{ij}^\dagger u_{ij})+(\psi_i^\dagger u_{ij}\psi_j+h.c.))\nonumber\\
&-&\sum_{<ij>}t_{ij}(b_i^\dagger u_{ij}b_j+h.c.)\label{eq:51}
\eeq
where $J$'s and $t$'s are determined by Eq. \ref{eq:201} and Eq. \ref{eq:202}, $u_{ij}$ is the mean field ansatz\cite{Lee}
\beq
u_{ij}=\delta_{\alpha\beta}\begin{pmatrix}
-\left<f_{i,\alpha}^\dagger f_{j,\beta}\right>^* & \left<f_{i,\alpha} f_{j,\beta}\right>\\
\left<f_{i,\alpha} f_{j,\beta}\right>^* & \left<f_{i,\alpha}^\dagger f_{j,\beta}\right>
\end{pmatrix}.
\eeq 
There are two kinds of constraints\cite{Lee}, namely proper filling
\beq
\left<b_i^\dagger b_i\right>= 2\delta
\eeq
and physical states
\beq
\left<\psi_i^\dagger\tau^l \psi_i +b_i^\dagger\tau^l b_i\right>=0,\quad l=1,2,3.
\eeq
where $\tau$'s are Pauli matrices. To implement these constraints, an additional penalty term should be added to the original Hamiltonian:
\beq
H_{\text{penalty}}=\sum_{i,l} {P( \left< \psi_i^\dagger\tau^l \psi_i +b_i^\dagger\tau^l b_i \right>)} \nonumber \\
-P(\sum_i \left< b_i^\dagger b_i \right>-2\delta),
\eeq
where $P(x)$ is a penalty function, $P(x)=G x^2$ with $G$ being huge (larger than $10^9$ in practice). In the determination of mean field ansatzes, it is required that $P(x)$ does not contribute to mean field energy in final solutions found.

\subsection{Projective Symmetry Groups Analysis and Schematic Phase Diagram}

Different phases characterized by mean field ansatzes of different kinds of $Z_2$ spin liquid are classified by PSG's\cite{Wen_2002}. The PSG's compose certain constraints on ansatzes, and at most 311 gauge inequivalent ansatzes are found. Details are presented in Appendix A. To further determine the phase diagram, the differential evolution algorithm in employed. With constraints from PSG's, number of optimizing variables are restricted to be 12, so that this global optimizing algorithm is sufficient.

\begin{figure}[htbp]
\centering
\includegraphics[width=8cm]{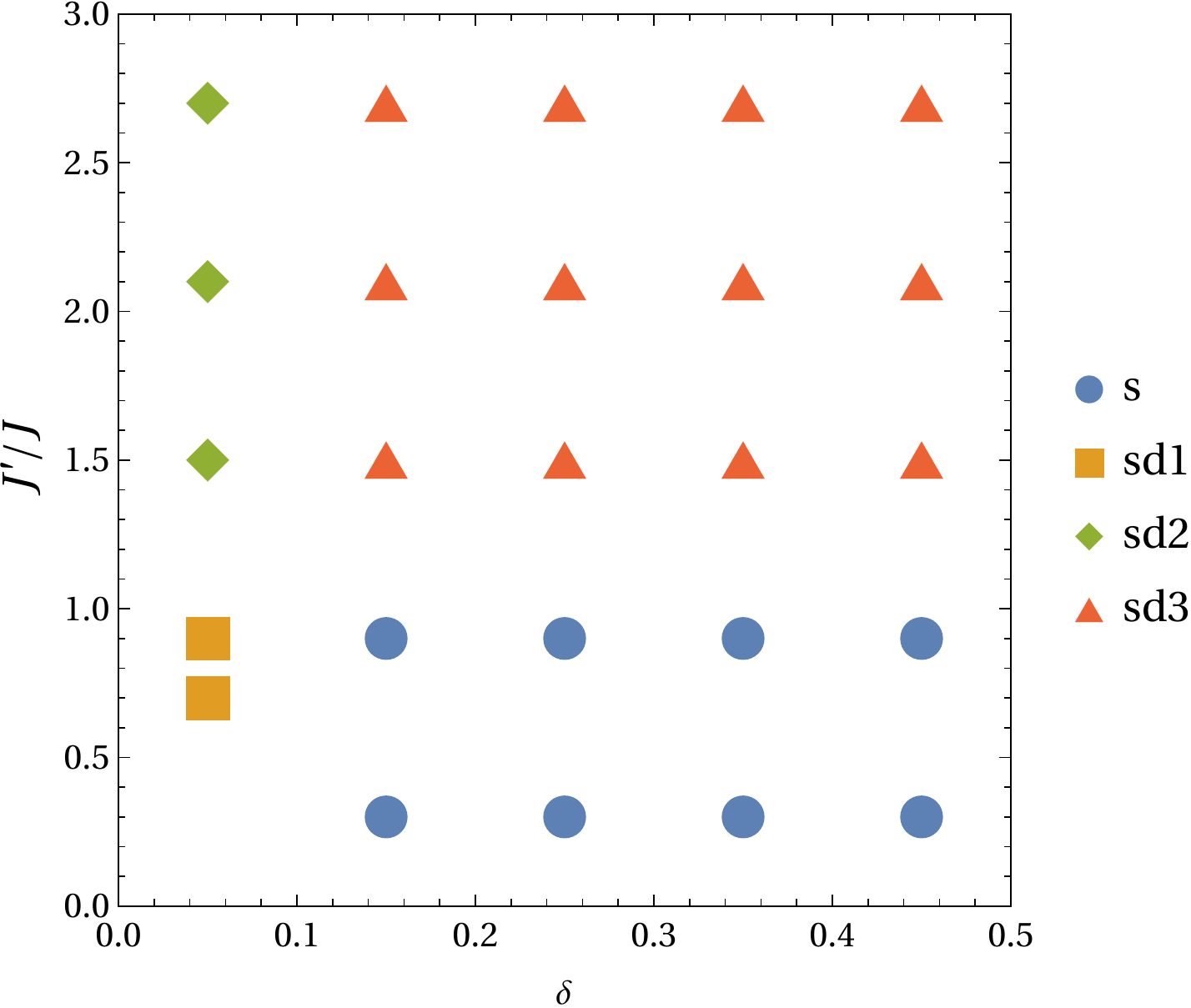}
\caption{Schematic phase diagram in strong coupling limit with $t=2t'=2t_y=0.04J_y=0.02J$. "s"and "sd" means $s_\pm^d$-wave and $(s+d)$-wave superconductivity, respectively. The bottom left corner of the phase diagram is left blank since numerical results are not reliable in this region.}\label{fig:7}
\end{figure}

$J'/J$ and filling $\delta$ are used as variables of the phase diagram. In what follows, the phase diagram is calculated in the case $t=2t'=2t_y=0.02J$ and $J_y=0.5J$. This set of coefficients satisfies that inter-chain hopping and coupling are smaller than intra-chain ones\cite{Liu}. The practical calculation is performed on a $16 \times 16 \times 16$ lattice with periodic boundary condition. For numerical details refer to Appendix B. With 25 choices of parameters investigated, the schematic phase diagram is obtained, as shown in Fig. \ref{fig:7}. However, in region $J'/J<0.5,\delta<0.1$, the numerical results are not reliable. This region of phase diagram is left blank.

In the region of coefficients we choose, there are 4 phases in total. At zero temperature, the holons necessarily condense, and the system is consequently in superconducting phases.

(i) The $s_\pm^d$-wave superconducting phase, namely $s_\pm$-wave with weak $d$-wave components, shown as "s" in Fig. \ref{fig:7}, described by
\beq
u_{i,i+\hat{x}}&=&\Delta_x \tau^1 -\chi_x \tau^3,\nonumber\\
u_{i,i+\hat{y}}&=&\Delta_y \tau^1 -\chi_y \tau^3,\nonumber\\
u_{i,i+\hat{z}}&=&u_{i,i+\hat{z}-\hat{x}}=u_{i,i+\hat{z}-\hat{y}}=u_{i,i+\hat{z}-\hat{x}-\hat{y}}\nonumber\\
&=&\Delta_z \tau^1 -\chi_z \tau^3 +i\chi_z' \tau^0.
\eeq

\begin{figure}[htbp]
\centering
\subfigure[]
{\includegraphics[width=4cm]{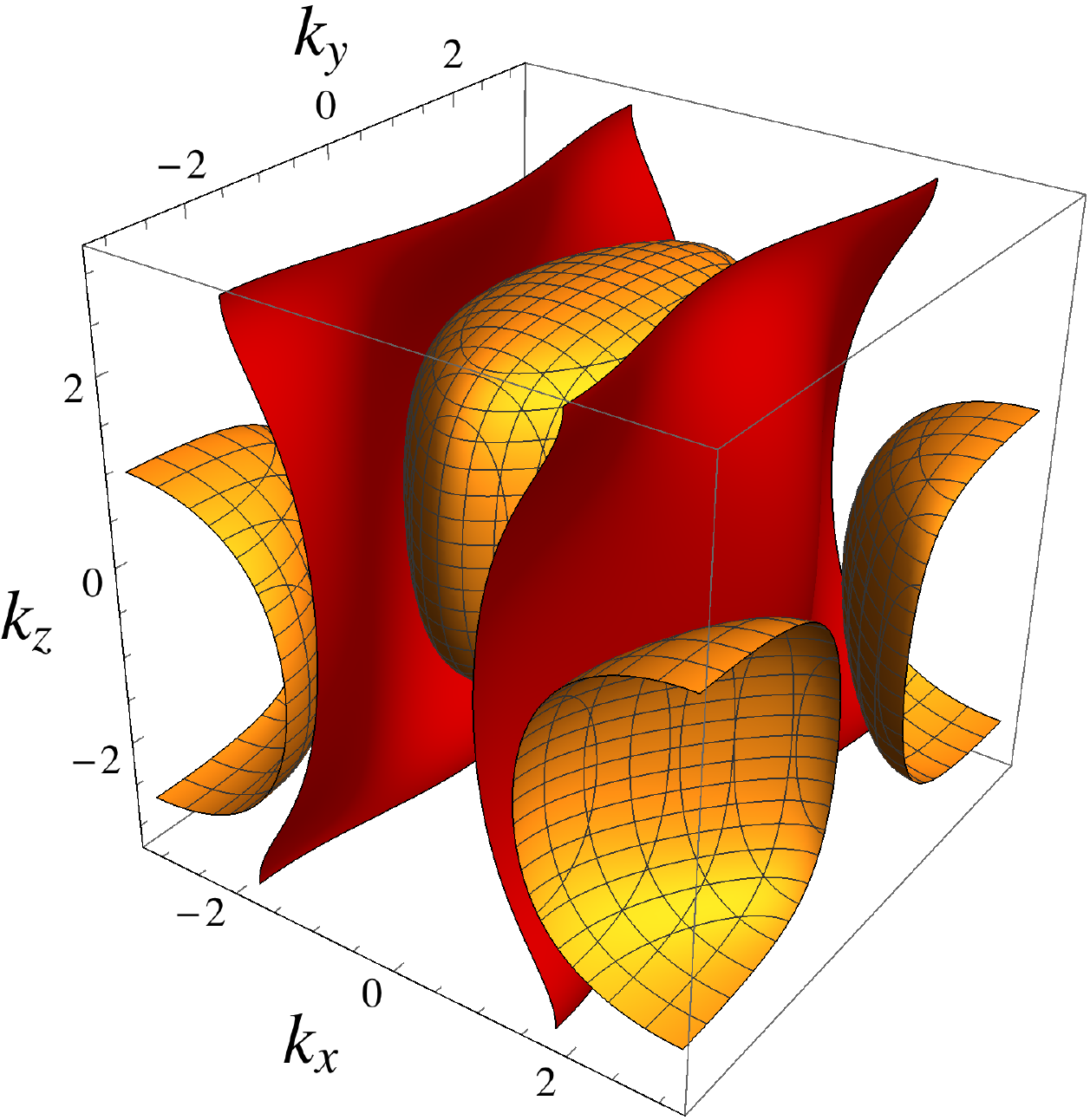}}
\subfigure[]
{\includegraphics[width=4cm]{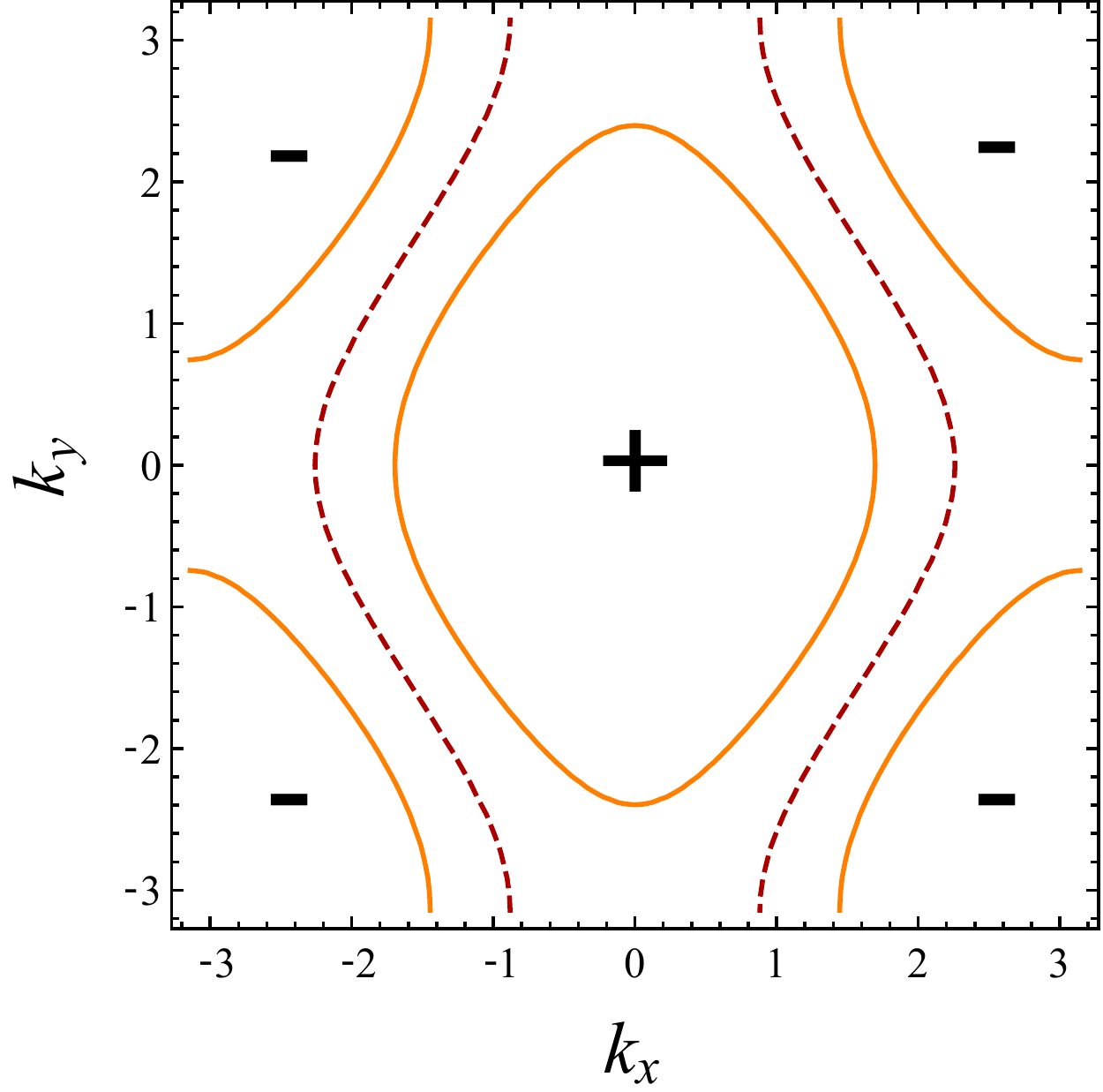}}
\caption{(a) Fermi surfaces shown by yellow meshed surfaces and zeros of superconducting gap shown by red meshless surfaces in $s_\pm^d$-wave phase. (b) The projection to $k_z=0$ plane, where Fermi surfaces are shown by yellow solid curves and zeros of superconducting gap are shown by red dashed curve. The sign of superconducting gap on Fermi surfaces are shown as "$+$" and "$-$". Here $\delta$=0.25.}\label{swave}
\end{figure}

Unlike what discussed in Ref.\cite{Wen_2002}, due to the absence of 4-fold rotation symmetry, weak $d$-wave components inevitably coexist in the $s$-wave superconducting phase. However, if the system exhibits $s$-wave superconductivity in general, it is still considered as an $s$-wave phase. The spinon Fermi surfaces and the zeros of superconducting gap of the $s_\pm^d$-wave phase we find are shown in Fig. \ref{swave}. Since holons condense in this phase, the spinon Fermi surfaces are the same as the electron Fermi surfaces. The superconducting gap on the Fermi surfaces has no node, which corroborates that the in this phase $s_\pm$-wave pairing is dominant.

(ii) The first $(s+d)$-wave superconducting phase, shown as "sd1" in Fig. \ref{fig:7}, described by
\beq
u_{i,i+\hat{x}}&=&\Delta_x \tau^1 -\chi_x \tau^3,\nonumber\\
u_{i,i+\hat{y}}&=&\Delta_y \tau^1 +\chi_y \tau^3,\nonumber\\
u_{i,i+\hat{z}}&=&u_{i,i+\hat{z}-\hat{x}}=u_{i,i+\hat{z}-\hat{y}}=u_{i,i+\hat{z}-\hat{x}-\hat{y}}\nonumber\\
&=&-\Delta_z \tau^1 -\chi_z \tau^3.
\eeq

(iii) The second $(s+d)$-wave superconducting phase, shown as "sd2" in Fig. \ref{fig:7}, described by
\beq
u_{i,i+\hat{x}}&=&\Delta_x \tau^1 -\chi_x \tau^3,\nonumber\\
u_{i,i+\hat{y}}&=&\Delta_y \tau^1 -\chi_y \tau^3,\nonumber\\
u_{i,i+\hat{z}}&=&-u_{i,i+\hat{z}-\hat{x}}=u_{i,i+\hat{z}-\hat{y}}=u_{i,i+\hat{z}-\hat{x}-\hat{y}}\nonumber\\
&=&\Delta_z \tau^1 -i\chi_z' \tau^0.
\eeq

(iv) The third $(s+d)$-wave superconducting phase, shown as "sd3" in Fig. \ref{fig:7}, described by
\beq
u_{i,i+\hat{x}}&=&\Delta_x \tau^1 -\chi_x \tau^3,\nonumber\\
u_{i,i+\hat{y}}&=&\Delta_y \tau^1 -\chi_y \tau^3,\nonumber\\
u_{i,i+\hat{z}}&=&-u_{i,i+\hat{z}-\hat{x}}=u_{i,i+\hat{z}-\hat{y}}=u_{i,i+\hat{z}-\hat{x}-\hat{y}}\nonumber\\
&=&\Delta_z \tau^1 -\chi_z \tau^3 -i\chi_z' \tau^0.
\eeq

As a comparison, we also find that the superconducting gap has nodes on Fermi surfaces in $(s+d)$-wave phases. All of the four ansatzes are consistent with the PSG analysis.

In conclusion, in strong coupling limit, the phase diagram is largely filled with the nodeless $s_\pm^d$-wave superconducting phase in the physical relevant regime of coefficients ($J'/J<1$).

\section{Weak Coupling Limit}

In the weak coupling limit ($t\gg J$), renormalization group (RG) and bosonization analysis are employed to determine possible phases.

\subsection{Quasi-1D Model}

\begin{figure}[htbp]
\centering
\includegraphics[width=0.95\linewidth]{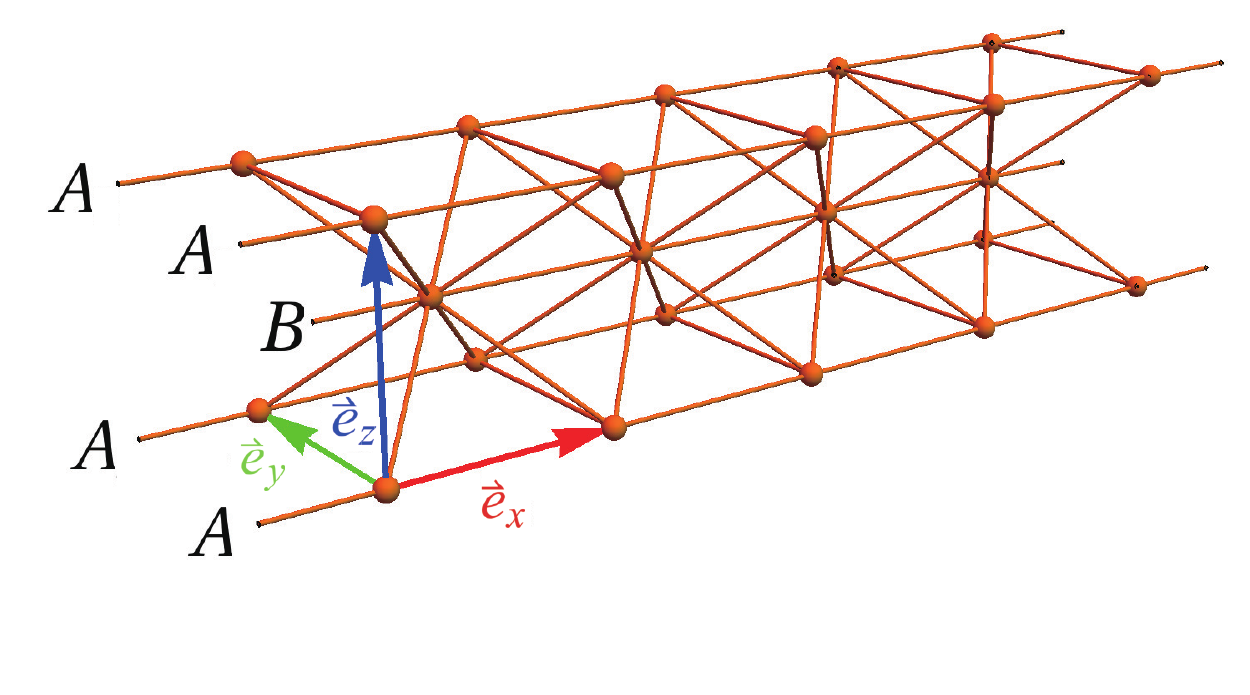}
\caption{The $N=2$ chains model. Here four $A$ chains are equivalent under periodic boundary condition, while $B$ chain is inequivalent to them. The unit cell is modified to be the conventional unit cell with two inequivalent atoms in one unit cell, and the $z$ direction is redefined.}\label{fig:3}
\end{figure}

\begin{figure}[htbp]
\centering
\includegraphics[width=6cm]{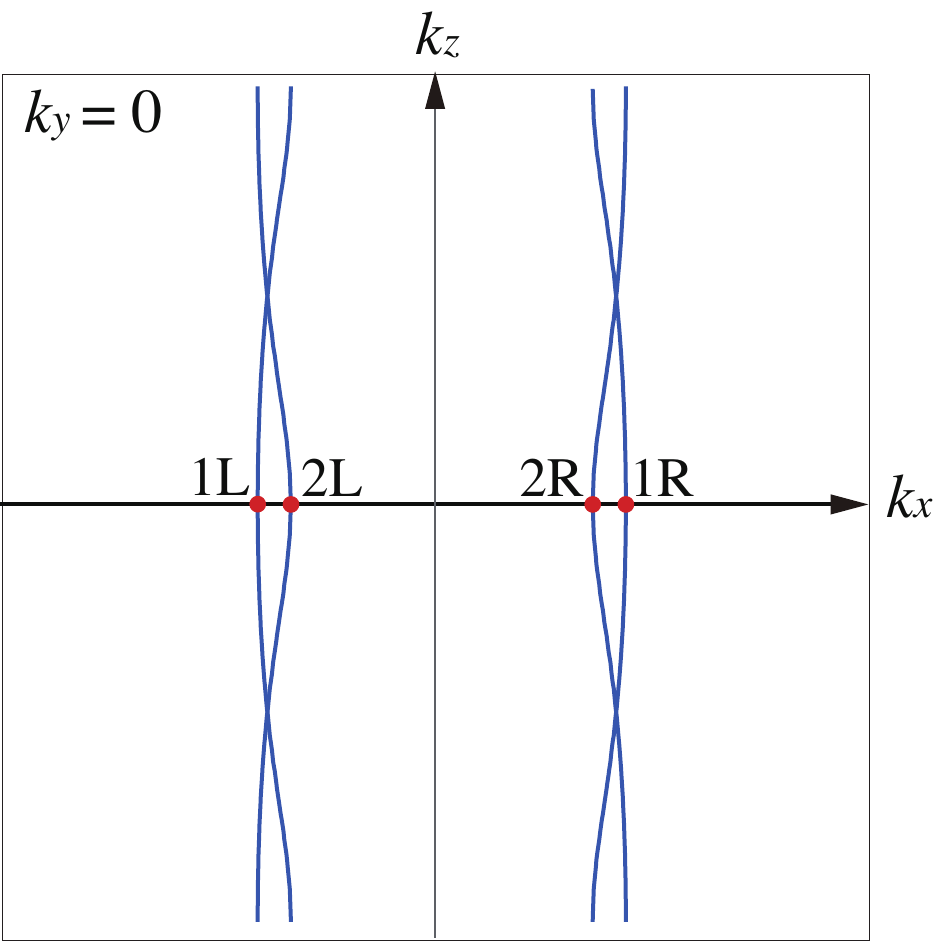}
\caption{The Fermi points named 1L, 1R and 2L, 2R as the intersection of the Fermi surfaces (blue curves) and line $k_y=k_z=0$. Here $\delta=0.2$.}\label{fig:4}
\end{figure}

An $N=2$ chains\cite{Lin} model is considered, as shown in Fig. \ref{fig:3}. Some previous work have focused on Luttinger liquids on two-leg ladders with\cite{Tonegawa_2017,Luo_2018,Giri:2017aa} or without frustration\cite{Scalapino:1998aa,Wu:2003aa,Lin,Balents_1996,Shelton:1998aa,Orignac:1997aa,Schulz:1996aa,Fabrizio:1993aa,Varma:1985aa}. However, the lattice structure in this model has not been investigated. Since the intra-chain coupling plays a more important role than that of the inter-chain coupling\cite{Liu}, and the translation symmetries of the conventional unit cells are presumably not destroyed, this quasi-1D model is believed to capture the most relevant physics. The unit cell is changed to be the conventional unit cell with two atoms in one unit cell, and the $z$ direction is redefined.
In the redefined coordinate, the non-interacting Hamiltonian $H_0$ can be diagonalized as
\beq
H_0=\sum_{\vec{k},s;i=1,2}\epsilon_i(\vec{k})\psi_{i,s}^{\dag}(\vec{k})\psi_{i,s}(\vec{k}),
\eeq
where
\beq
\epsilon_{1,2}(\vec{k})=-2(&\pm &t'(\cos(k_x+k_y+\frac{k_z}{2})+\cos(k_x-k_y+\frac{k_z}{2})\nonumber\\
&+&\cos(-k_x+k_y+\frac{k_z}{2})+\cos(k_x-k_y-\frac{k_z}{2}))\nonumber\\
&+&t\cos(k_x)+t_y\cos(k_y)),
\eeq
where $+(-)$ sign for $\epsilon_{1(2)}$, respectively. For the $N=2$ chains model, the summation over $\vec{k}$ only contains those points with $k_y=k_z=0$. Therefore, the Fermi points are determined via\cite{Lin}
\beq
\epsilon_i(k_{Fi})=\mu,\quad i=1,2,
\eeq
for chemical potential $\mu$. The Fermi points in this quasi-1D model can be viewed as a discrete set of points with $k_y=k_z=0$ on the 3D Fermi surface of the $N=+\infty$ model. As shown in Fig. \ref{fig:4}. For clarity the $k_y$ direction is neglected. For a generic filling, the Fermi points does not coincide, and there is no umklapp interactions.

Only excitations around Fermi points are considered in long wave length limit. Field operators can be written in terms of chiral fermions (right/left movers) as\cite{Lin}
\beq
\psi_{i,s}\sim\psi_{i,s}^Re^{ik_{Fi}x}+\psi_{i,s}^Le^{-ik_{Fi}x}, \quad i=1,2,
\eeq 
with $\psi_{1,s}^{L}$, $\psi_{1,s}^{R}$, $\psi_{2,s}^{L}$ and $\psi_{2,s}^{R}$ corresponding to excitation around Fermi point 1L, 1R, 2L and 2R, respectively. The momenta of these chiral fermions are bounded by a momentum cut-off $\Lambda\ll k_{F1,2}$. Therefore, the dispersion can be linearized within $\Lambda$. The effective non-interacting Hamiltonian reads
\beq
H_0=\sum_{i,s}\int dxv_i(\psi_{i,s}^{R\dag}i\partial_x\psi_{i,s}^R-\psi_{i,s}^{L\dag}i\partial_x\psi_{i,s}^L),
\eeq
where $v_i=\partial_{k_x}\epsilon_i(\vec{k})\mid_{k_x=k_{Fi},k_y=k_z=0}$ is the Fermi velocity.

\subsection{Renormalization Group and Bosonization Analysis}

A generic interaction Hamiltonian density subject to the constraint of momenta conservation reads
\beq
\mathcal{H'}&=&\sum_{i,j=1}^2f_{ij}^{\rho}T_{ii}^R T_{jj}^L-f_{ij}^{\sigma}\vec{T}_{ii}^R\cdot\vec{T}_{jj}^L\nonumber\\
&+&\sum_{i=1}^2{f'}_{ii}^{\rho}T_{i,3-i}^RT_{i,3-i}^L-{f'}_{i,i}^{\sigma}\vec{T}_{i,3-i}^R\cdot \vec{T}_{i,3-i}^L,
\eeq
where currents
\beq
T_{ij}&=&\sum_{s}\psi_{i,s}^{\dag}\psi_{j,s},\\
\vec{T}_{ij}&=&\frac{1}{2}\sum_{s,s'}\psi_{i,s}^{\dag}\vec{\sigma}_{ss'}\psi_{j,s'}.
\eeq
Coupling constants $f$'s and ($f'$)'s represent intra-band and inter-band scattering, respectively. The relationship of their values are given by certain symmetries. Charge conjugation $T_{ij}\rightarrow T_{ji}$ gives ${f'}_{ii}={f'}_{3-i,3-i}$, while reflection in $x$ direction gives\cite{Lin} $f_{ij}=f_{ji}$. Details of the construction of this interaction Hamiltonian density is left in Appendix C.

To construct a low energy effective theory, the interaction is renormalized and bosonized. The derivation of the RG equations is left in Appendix D. After solving RG equations numerically, in the region of coefficients adopted, we find that in all cases there are two coupling constants, $(f_{11}^\rho,f_{11}^\sigma)$ or $(f_{22}^\rho,f_{22}^\sigma)$, dominant. Since $f^\rho$ only contribute gradient term after bosonization\cite{Lin}, they are simply dropped. Therefore, the interaction Hamiltonian after RG reads (take subscript 1 as example)
\beq
\mathcal{H}'=\sum_{s}\frac{1}{2} f^\sigma_{11}\psi_{1,s}^{R\dag}\psi_{1,\bar{s}}^{L\dag}\psi_{1,\bar{s}}^{L}\psi_{1,s}^{R},\label{eq:501}
\eeq
where $\bar{s}$ means the opposite direction of spin $s$. After bosonization, in terms of the chiral boson fields, the Hamiltonian reads (take subscript 1 as an example)
\beq
\mathcal{H}_0&\sim &\frac{v_1}{2}((\partial_x \theta_{1,\sigma})^2+(\partial_\tau \theta_{1,\sigma})^2),\\
\mathcal{H}'&\sim & f^\sigma_{11}\cos(\sqrt{8\pi}\theta_{1,\sigma}),\label{eq:101}
\eeq
Purely free fields are neglected in $\mathcal{H}_0$. Therefore, the low energy effective theory of the system is a sine-Gordon theory. The bosonization dictionary is left in Appendix E, including the definition of $\theta_{i,\sigma}$.

\subsection{Phase Diagram}

The global minima of Eq. \ref{eq:101} is
\beq
\sqrt{2\pi}\theta_{i,\sigma}=2l_i\pi \quad \text{or}\quad(2l_i+1)\pi, \quad l_i\in \mathbb{Z},
\eeq
depending on the sign of $f^{\sigma}_{ii}$. Around a minimum, the interaction Hamiltonian can be expanded as $\mathcal{H}'\sim m(\delta\theta)^2$, which gives the field $\theta$ an effective mass. Therefore, when one $f^\sigma$ is dominant, there is one gapless spin mode and one gapped spin mode. The two charge modes are always gapless. To figure out the phase diagrams, the expectation values of different order parameters are calculated, including charge density wave (CDW), spin density wave (SDW), singlet superconductivity (SS) and triplet superconductivity (TS):\cite{Fradkin}
\beq
\mathcal{O}_{\text{CDW},i}&=&\sum_s \psi_{i,s}^{R\dag}(x)\psi_{i,s}^{L}(0),\\
\vec{\mathcal{O}}_{\text{SDW},i}&=&\sum_{s,s'} \psi_{i,s}^{R\dag}(x)\vec{\sigma}_{ss'}\psi_{i,s'}^{L}(0),\\
\mathcal{O}_{\text{SS},i}&=&\psi_{i,\uparrow}^{R\dag}(x)\psi_{i,\downarrow}^{L\dag}(0),\\
\mathcal{O}_{\text{TS},i}&=&\psi_{i,\uparrow}^{R\dag}(x)\psi_{i,\uparrow}^{L\dag}(0).
\eeq
These order parameters can be rewritten in terms of boson fields via the bosonization dictionary in Appendix E. As indicated in Ref.\cite{Lin}, according to the uncertainty principle $[\phi,\theta]=O(1)$, the conjugate field of $\theta_\sigma$, namely $\phi_\sigma$, fluctuates violently. Therefore, only terms like $e^{\phi_\sigma(x)-\phi_\sigma(0)}$ can survive in the mean field level. Applying this criteria, one can determine whether the order parameters are non-vanishing in certain phases.

Without losing generality, $f^\sigma_{11}$ is supposed to be dominant. All the non-vanishing order parameters are $\mathcal{O}_{\text{CDW}}$, $\vec{\mathcal{O}}_{\text{SDW}}$ and $\mathcal{O}_{\text{SS}}$. When $f^\sigma_{11}$ is negative, according to the Cooper instability, an attractive interaction will naturally induce superconductivity. Therefore, the system will be in an $s$-wave superconducting phase (with weak $d$-wave components due to the absence of four-fold rotation symmetry). When $f^\sigma_{11}$ is positive, both CDW and SDW can exist in this system, due to the gaplessness of the charge modes and the spin mode. The system will present spin-charge separation and hence in a Luttinger liquid phase with two gapless charge modes and one gapless spin mode.

\begin{figure}[htbp]
\centering
\includegraphics[width=8cm]{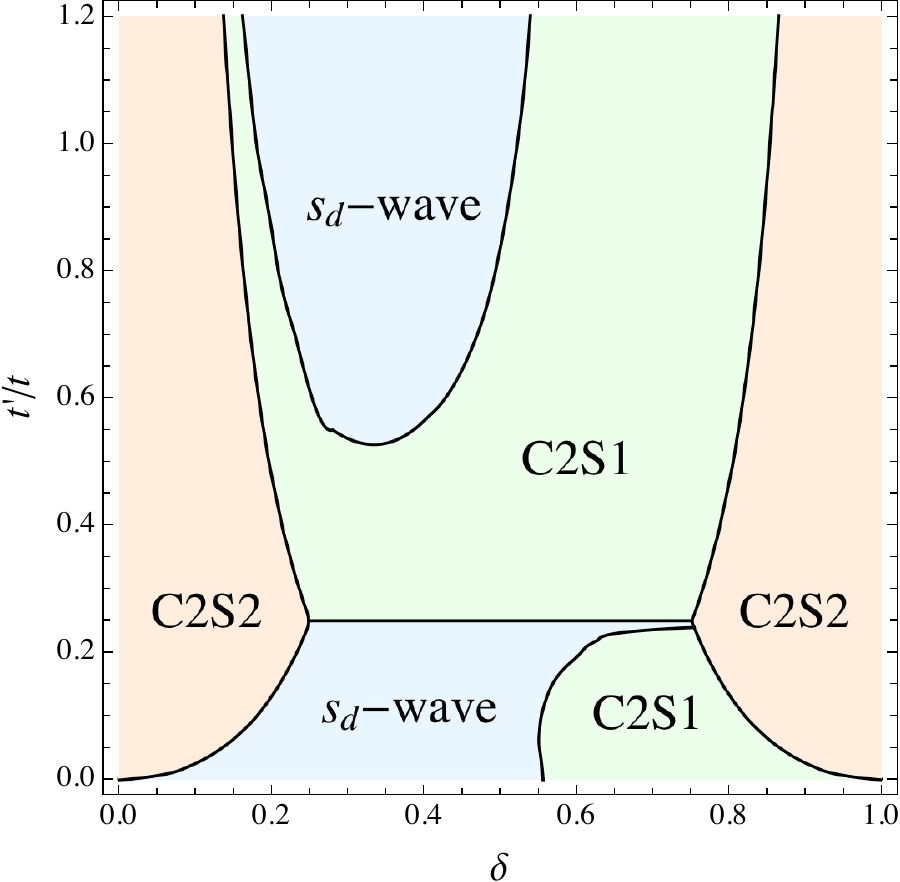}
\caption{Phase diagram in weak coupling limit with $J=3J'=6J_y=0.06t_y=0.03t$. "C$m$S$n$" stands for Luttinger liquid phase with $m$ gapless charge modes and $n$ gapless spin modes\cite{Lin}. "$s_d$-wave" stands for $s$-wave superconducting phase with weak $d$-wave components.}\label{fig:5}
\end{figure}

The phase diagram is shown in Fig. \ref{fig:5}. We use $t'/t$ and filling $\delta$ as variables, and fix other coefficients as $t_y=0.5t$ and $J=3J'=6J_y=0.03t$ to satisfy that inter-chain hopping and coupling are smaller than intra-chain ones\cite{Liu}. The phase diagram is largely occupied by $s$-wave superconducting phase with weak $d$-wave components (denoted as $s_d$-wave in Fig. \ref{fig:5}). For appropriate filling, one of the two bands is fully empty or fully occupied, leading to only two Fermi points, instead of four, participating in interactions. In this case, the only dominant coupling constants are $f^\rho$'s, implying all of the two charge modes and two spin modes are gapless in this Luttinger liquid phase denoted as\cite{Lin} "C2S2" in Fig. \ref{fig:5}. The other Luttinger liquid phase with two gapless charge modes and one spin mode is denoted as\cite{Lin} "C2S1". In the $s_d$-wave phase, the pairing order parameter has no node in $k$-space, since the mean field decomposition of interaction Hamiltonian density Eq. \ref{eq:501} can be rearranged into the form of a traditional BCS Hamiltonian. Therefore, the superconducting gap has the same sign on all the four Fermi points, 1L, 1R and 2L, 2R. The superconducting phase is therefore a nodeless $s_d$-wave phase.

\section{Conclusions}

In this work zero-temperature phases of cuprates with Ba$_2$CuO$_{3+\delta}$-type CuO chain structure are investigated in both strong and weak coupling limits of a single-orbital multi-chain t-J model. We find that in both of the two limits, the phase diagrams are largely filled with nodeless $s$-wave superconducting phases (with weak $d$-wave components). It is $s_\pm$-wave with weak $d$-wave components (denoted as $s^d_\pm$-wave) in strong coupling limit, and $s$-wave with weak $d$-wave components (denoted as $s_d$-wave) in weak coupling limit. We believe that our conclusion of nodeless $s$-wave pairing in Ba$_2$CuO$_{3+\delta}$ is consistent with the experimental observations of the stability of high-$T_C$ against disorder\cite{Li:2019aa,PhysRevLett.3.325,Balatsky:2006aa}. Some previous theories\cite{Mai:2019,Ya:2020} also proposed $s_\pm$-wave phases. However, they are based on multi-orbital models and developed on lattice structures apparently different from ours. $d$-wave superconductivity was also proposed in previous works\cite{Wang:2020,Le:2019}. Our proposed nodeless $s$-wave pairing symmetry can in principle be detected in phase sensitive\cite{Harlingen:1995aa,Tsuei:2000aa} and spectroscopic\cite{Damascelli:2003aa} measurements.

\section{Acknowledgement}

ZQG and KWS acknowledge Hui Yang, Jie-Ran Xue and Xue-Mei Wang for enlightening discussions. FW acknowledges support from The National Key Research and Development Program of China (Grant No. 2017YFA0302904), and National Natural Science Foundation of China (Grant No. 11888101).

\appendix

\section{Classification of Z$_2$ Spin Liquids Phases in Ba$_2$CuO$_{3+\delta}$}

\subsection{Projective Symmetry Groups}

Under coordinates we choose in Sec. III, space group symmetries, including translation, parity, and inversion, read
\beq
T_x:(i_x,i_y,i_z)&\mapsto &(i_x+1,i_y,i_z),\\
T_y:(i_x,i_y,i_z)&\mapsto &(i_x,i_y+1,i_z),\\
T_z:(i_x,i_y,i_z)&\mapsto &(i_x,i_y,i_z+1),\\
P_x:(i_x,i_y,i_z)&\mapsto &(-i_x-i_z,i_y,i_z),\\
P_y:(i_x,i_y,i_z)&\mapsto &(i_x,-i_y-i_z,i_z),\\
I:(i_x,i_y,i_z)&\mapsto &(-i_x,-i_y,-i_z),
\eeq
and the time-reversal symmetry is
\beq
\mathcal{T}:u_{ij}\mapsto -u_{ij}.
\eeq
The symmetries above satisfy equalities
\beq
T_x^{-1}T_y^{-1}T_xT_y&=&1\label{eq:1},\\
T_y^{-1}T_z^{-1}T_yT_z&=&1\label{eq:2},\\
T_z^{-1}T_x^{-1}T_zT_x&=&1\label{eq:3},\\
T_xP_xT_xP_x&=&1\label{eq:4},\\
T_yP_yT_yP_y&=&1,\\
T_x^{-1}P_yT_xP_y&=&1,\\
T_y^{-1}P_xT_yP_x&=&1,\\
P_xT_xT_z^{-1}P_xT_z&=&1,\\
P_yT_yT_z^{-1}P_yT_z&=&1,\\
P_xP_yP_xP_y&=&1,\\
P_x^2=P_y^2&=&1\label{eq:5},\\
T_xIT_xI&=&1\label{eq:10},\\
T_yIT_yI&=&1,\\
T_zIT_zI&=&1\label{eq:11},\\
P_xIP_xI&=&1\label{eq:12},\\
P_yIP_yI&=&1\label{eq:13},\\
I^2&=&1.
\eeq
$\mathcal{T}$ commutes with all the space group symmetries. Following Ref\cite{Wen_2002}, we can first determine $G_x$, $G_y$, $G_z$, and $G_T$ through Eq.\ref{eq:1}, Eq.\ref{eq:2} Eq.\ref{eq:3} and the commutation relations between $\mathcal{T}$ and translations. Unlike the 2D case\cite{Wen_2002}, here we choose the gauge that
\beq
G_z(i)&=&\tau ^0,\\
G_x(i)&=&\eta _x^{i_z}\tau ^0,\\
G_y(i)&=&\eta _y^{i_z}\eta _{xy}^{i_x}\tau ^0,\\
G_T(i)&=&g_T\eta_{xt}^{i_x}\eta_{yt}^{i_y}\eta_{zt}^{i_z}\label{eq:18},
\eeq
where two gauge inequivalent choices of $g_T$ are $g_T=\tau^0$ or $i\tau^3$, and seven $\eta$s are $\pm1$.

Then we consider parities $P_x$ and $P_y$. The PSG equations given by Eq.\ref{eq:4} to Eq.\ref{eq:5} read
\beq
G_x(P_x(i))G^{-1}_{P_x}(i+\hat{x})G_x(i+\hat{x})G_{P_x}(i)&\in &\mathcal{G}\label{eq:6},\\
G_y^{-1}(P_x(i))G^{-1}_{P_x}(i)G_y(i)G_{P_x}(i-\hat{y})&\in &\mathcal{G},\\
G_x^{-1}(P_y(i))G^{-1}_{P_y}(i)G_x(i)G_{P_y}(i-\hat{x})&\in &\mathcal{G},\\
G_y(P_y(i))G^{-1}_{P_y}(i+\hat{y})G_y(i+\hat{y})G_{P_y}(i)&\in &\mathcal{G}\label{eq:7},\\
G_{P_x}(i)G_x(P_x(i))G^{-1}_z(T_zT_x^{-1}P_x(i))\cdot \nonumber \\
G_{P_x}(T_zT_x^{-1}P_x(i))G_z(T_zT_x^{-1}(i))&\in &\mathcal{G}\label{eq:8},\\
G_{P_y}(i)G_y(P_y(i))G^{-1}_z(T_zT_y^{-1}P_y(i))\cdot \nonumber \\
G_{P_y}(T_zT_y^{-1}P_y(i))G_z(T_zT_y^{-1}(i))&\in &\mathcal{G},\\
G_{P_x}(i)G_{P_y}(P_x(i))G^{-1}_{P_x}(P_y(i))G^{-1}_{P_y}(i)&\in &\mathcal{G},\\
G_{P_x}(i)G_{P_x}(P_x(i))&\in &\mathcal{G},\\
G_{P_y}(i)G_{P_y}(P_y(i))&\in &\mathcal{G}\label{eq:9}.
\eeq
Since $P_x$ and $P_y$ do not change $i_z$, in our gauge Eq.\ref{eq:6} to Eq.\ref{eq:7} reduce to
\beq
G^{-1}_{P_x}(i+\hat{x})G_{P_x}(i)&=&\eta_{xpx}\tau ^0,\\
G^{-1}_{P_x}(i+\hat{y})G_{P_x}(i)&=&\eta_{ypx}\eta_{xy}^{i_z}\tau ^0,\\
G^{-1}_{P_y}(i+\hat{x})G_{P_y}(i)&=&\eta_{xpy}\tau ^0,\\
G^{-1}_{P_y}(i+\hat{y})G_{P_y}(i)&=&\eta_{ypy}\tau ^0,
\eeq
which give $G_{P_x}(i)$ and $G_{P_y}(i)$ the generic form
\beq
G_{P_x}(i)&=&g_{P_x}\Theta _{P_x}(i_z)\eta_{xpx}^{i_x}\eta_{ypx}^{i_y}\label{eq:19},\\
G_{P_y}(i)&=&g_{P_y}\Theta _{P_y}(i_z)\eta_{xpy}^{i_x}\eta_{ypy}^{i_y}\eta_{xy}^{i_yi_z}\label{eq:20},
\eeq
where $\Theta$s are ($\pm1$)-valued functions of $i_z$. Eq.\ref{eq:8} to Eq.\ref{eq:9} then reduce to 
\beq
g_{P_x}^2\Theta _{P_x}(i_z)\Theta _{P_x}(i_z+1)\eta_{xpx}^{i_z+1}G_x(i)&=&\eta_{5px} \tau ^0\label{eq:16},\\
g_{P_y}^2\Theta _{P_y}(i_z)\Theta _{P_y}(i_z+1)\eta_{xy}^{i_y}G_y(i)&=&\eta_{5py} \tau ^0\label{eq:17},\\
g_{P_x}g_{P_y}g^{-1}_{P_x}g^{-1}_{P_y}(\eta_{xpy}\eta_{ypx})^{i_z}&=&\pm \tau ^0\label{eq:14},\\
g^2_{P_x}\eta_{xpx}^{i_z}&=&\pm \tau ^0,\\
g^2_{P_y}(\eta_{ypy}\eta_{xy})^{i_z}&=&\pm \tau ^0\label{eq:15},
\eeq
for all sites $i$. Eq.\ref{eq:14} to Eq.\ref{eq:15} require that
\beq
\eta_{xpy}=\eta_{ypx},\eta_{xy}=\eta_{xpx}=\eta_{ypy}=1,
\eeq
while Eq.\ref{eq:16} and Eq.\ref{eq:17} give two $\Theta$s a specific form. All gauge inequivalent $\Theta$s are
\beq
\Theta _{P_x}(i_z)&=&\eta_{5px}^{i_z},\text{ for }G_x(i)=\tau ^0;\\
\Theta _{P_y}(i_z)&=&\eta_{5py}^{i_z},\text{ for }G_y(i)=\tau ^0;\\
\Theta _{P_x}(i_z)&=&\eta_{5px}^{i_z}\cdot\sqrt{2}\sin(\frac{\pi}{2}i_z+\frac{\pi}{4}),\nonumber\\ \text{for }G_x(i)&=&(-)^{i_z}\tau ^0;\\
\Theta _{P_y}(i_z)&=&\eta_{5py}^{i_z}\cdot\sqrt{2}\sin(\frac{\pi}{2}i_z+\frac{\pi}{4}),\nonumber\\ \text{for }G_y(i)&=&(-)^{i_z}\tau ^0.
\eeq

Finally we consider inversion $I$. Eq.\ref{eq:10} to Eq.\ref{eq:11} induce PSG equations
\beq
G_x(I(i))G_{I}(i+\hat{x})G_x(i+\hat{x})G_{I}(i)&\in & \mathcal{G},\\
G_y(I(i))G_{I}(i+\hat{y})G_y(i+\hat{y})G_{I}(i)&\in &\mathcal{G},\\
G_z(I(i))G_{I}(i+\hat{z})G_z(i+\hat{z})G_{I}(i)&\in &\mathcal{G}.
\eeq
Under our gauge, $G_{I}(i)$ has the generic form
\beq
G_{I}(i)=g_I\eta_{xI}^{i_x}\eta_{yI}^{i_y}\eta_{zI}^{i_z}\label{eq:21}.
\eeq
According to Eq.\ref{eq:12} and Eq.\ref{eq:13},
\beq
G_I(P_x(i))G^{-1}_{P_x}(I(i))G_x(I(i))G_{P_x}(i)&\in &\mathcal{G},\\
G_I(P_y(i))G^{-1}_{P_y}(I(i))G_y(I(i))G_{P_y}(i)&\in &\mathcal{G},
\eeq
we have
\beq
g_Ig^{-1}_{P_x}g_Ig_{P_x}\eta_{xI}^{i_z}&=&\pm \tau ^0,\\
g_Ig^{-1}_{P_y}g_Ig_{P_y}\eta_{yI}^{i_z}&=&\pm \tau ^0,
\eeq
for all sites $i$. Therefore, $\eta_{xI}=\eta_{yI}=1$. From just the same argument, $\eta_{xt}=\eta_{yt}=1$. Then, Eq.\ref{eq:18}, Eq.\ref{eq:19}, Eq.\ref{eq:20} and Eq.\ref{eq:21} reduce to
\beq
G_T(i)&=&g_T\eta_{t}^{i_z},\\
G_{P_x}(i)&=&g_{P_x}\Theta _{P_x}(i_z)\eta_{p}^{i_y},\\
G_{P_y}(i)&=&g_{P_y}\Theta _{P_y}(i_z)\eta_{p}^{i_x},\\
G_I(i)&=&g_I\eta_{I}^{i_z},
\eeq
where $\eta_{t}$, $\eta_{p}$ and $\eta_{I}$ can take value of $\pm1$, and $\Theta _{P_x}(i_z)$ and $\Theta _{P_y}(i_z)$ are determined above. The constraints of $g$s reduce to
\beq
g_{P_x}^2=g_{P_y}^2=g_I^2&=&\pm \tau ^0,\\
g_{P_x}g_{P_y}g^{-1}_{P_x}g^{-1}_{P_y}&=&\pm \tau ^0,\\
g_Ig^{-1}_{P_x}g_Ig_{P_x}&=&\pm \tau ^0,\\
g_Ig^{-1}_{P_y}g_Ig_{P_y}&=&\pm \tau ^0,\\
g_Tg^{-1}_{P_x}g_Tg_{P_x}&=&\pm \tau ^0,\\
g_Tg^{-1}_{P_y}g_Tg_{P_y}&=&\pm \tau ^0,\\
g_Ig^{-1}_Tg_Ig_T&=&\pm \tau ^0.
\eeq
All gauge inequivalent choices of $g$s are
\beq
g_T&=&\tau ^0\quad g_{P_x}=\tau ^0\quad g_{P_y}=\tau ^0\quad g_I=\tau ^0;\\
g_T&=&\tau ^0\quad g_{P_x}=i\tau ^3\quad g_{P_y}=i\tau ^3\quad g_I=i\tau ^3;
\eeq
\beq
g_T&=&i\tau ^3\quad g_{P_x}=\tau ^0\quad g_{P_y}=\tau ^0\quad g_I=\tau ^0;\\
g_T&=&i\tau ^3\quad g_{P_x}=i\tau ^3\quad g_{P_y}=i\tau ^3\quad g_I=i\tau ^3;\\
g_T&=&i\tau ^3\quad g_{P_x}=i\tau ^1\quad g_{P_y}=i\tau ^1\quad g_I=i\tau ^1;
\eeq
\beq
g_T&=&\tau ^0\quad g_{P_x}=i\tau ^3\quad g_{P_y}=\tau ^0\quad g_I=\tau ^0;\\
g_T&=&\tau ^0\quad g_{P_x}=\tau ^0\quad g_{P_y}=i\tau ^3\quad g_I=\tau ^0;\\
g_T&=&\tau ^0\quad g_{P_x}=\tau ^0\quad g_{P_y}=\tau ^0\quad g_I=i\tau ^3;
\eeq
\beq
g_T&=&i\tau ^3\quad g_{P_x}=i\tau ^3\quad g_{P_y}=\tau ^0\quad g_I=\tau ^0;\\
g_T&=&i\tau ^3\quad g_{P_x}=\tau ^0\quad g_{P_y}=i\tau ^3\quad g_I=\tau ^0;\\
g_T&=&i\tau ^3\quad g_{P_x}=\tau ^0\quad g_{P_y}=\tau ^0\quad g_I=i\tau ^3;\\
g_T&=&i\tau ^3\quad g_{P_x}=i\tau ^1\quad g_{P_y}=\tau ^0\quad g_I=\tau ^0;\\
g_T&=&i\tau ^3\quad g_{P_x}=\tau ^0\quad g_{P_y}=i\tau ^1\quad g_I=\tau ^0;\\
g_T&=&i\tau ^3\quad g_{P_x}=\tau ^0\quad g_{P_y}=\tau ^0\quad g_I=i\tau ^1;
\eeq
\beq
g_T&=&\tau ^0\quad g_{P_x}=i\tau ^3\quad g_{P_y}=i\tau ^3\quad g_I=\tau ^0;\\
g_T&=&\tau ^0\quad g_{P_x}=i\tau ^3\quad g_{P_y}=\tau ^0\quad g_I=i\tau ^3;\\
g_T&=&\tau ^0\quad g_{P_x}=\tau ^0\quad g_{P_y}=i\tau ^3\quad g_I=i\tau ^3;
\eeq
\beq
g_T&=&i\tau ^3\quad g_{P_x}=i\tau ^3\quad g_{P_y}=i\tau ^3\quad g_I=\tau ^0;\\
g_T&=&i\tau ^3\quad g_{P_x}=i\tau ^3\quad g_{P_y}=\tau ^0\quad g_I=i\tau ^3;\\
g_T&=&i\tau ^3\quad g_{P_x}=\tau ^0\quad g_{P_y}=i\tau ^3\quad g_I=i\tau ^3;\\
g_T&=&i\tau ^3\quad g_{P_x}=i\tau ^1\quad g_{P_y}=i\tau ^1\quad g_I=\tau ^0;\\
g_T&=&i\tau ^3\quad g_{P_x}=i\tau ^1\quad g_{P_y}=i\tau ^0\quad g_I=i\tau ^1;\\
g_T&=&i\tau ^3\quad g_{P_x}=\tau ^0\quad g_{P_y}=i\tau ^1\quad g_I=i\tau ^1;
\eeq
\beq
g_T&=&\tau ^0\quad g_{P_x}=i\tau ^3\quad g_{P_y}=i\tau ^1\quad g_I=\tau ^0;\\
g_T&=&\tau ^0\quad g_{P_x}=i\tau ^1\quad g_{P_y}=\tau ^0\quad g_I=i\tau ^3;\\
g_T&=&\tau ^0\quad g_{P_x}=\tau ^0\quad g_{P_y}=i\tau ^3\quad g_I=i\tau ^1;
\eeq
\beq
g_T&=&i\tau ^3\quad g_{P_x}=i\tau ^3\quad g_{P_y}=i\tau ^1\quad g_I=\tau ^0;\\
g_T&=&i\tau ^3\quad g_{P_x}=i\tau ^1\quad g_{P_y}=\tau ^0\quad g_I=i\tau ^3;\\
g_T&=&i\tau ^3\quad g_{P_x}=\tau ^0\quad g_{P_y}=i\tau ^3\quad g_I=i\tau ^1;\\
g_T&=&i\tau ^3\quad g_{P_x}=i\tau ^1\quad g_{P_y}=i\tau ^3\quad g_I=\tau ^0;\\
g_T&=&i\tau ^3\quad g_{P_x}=i\tau ^3\quad g_{P_y}=\tau ^0\quad g_I=i\tau ^1;\\
g_T&=&i\tau ^3\quad g_{P_x}=\tau ^0\quad g_{P_y}=i\tau ^1\quad g_I=i\tau ^3;
\eeq
\beq
g_T&=&\tau ^0\quad g_{P_x}=i\tau ^3\quad g_{P_y}=i\tau ^3\quad g_I=i\tau ^1;\\
g_T&=&\tau ^0\quad g_{P_x}=i\tau ^3\quad g_{P_y}=i\tau ^1\quad g_I=i\tau ^3;\\
g_T&=&\tau ^0\quad g_{P_x}=i\tau ^1\quad g_{P_y}=i\tau ^3\quad g_I=i\tau ^3;
\eeq
\beq
g_T&=&i\tau ^3\quad g_{P_x}=i\tau ^3\quad g_{P_y}=i\tau ^3\quad g_I=i\tau ^1;\\
g_T&=&i\tau ^3\quad g_{P_x}=i\tau ^3\quad g_{P_y}=i\tau ^1\quad g_I=i\tau ^3;\\
g_T&=&i\tau ^3\quad g_{P_x}=i\tau ^1\quad g_{P_y}=i\tau ^3\quad g_I=i\tau ^3;\\
g_T&=&i\tau ^3\quad g_{P_x}=i\tau ^3\quad g_{P_y}=i\tau ^1\quad g_I=i\tau ^1;\\
g_T&=&i\tau ^3\quad g_{P_x}=i\tau ^1\quad g_{P_y}=i\tau ^3\quad g_I=i\tau ^1;\\
g_T&=&i\tau ^3\quad g_{P_x}=i\tau ^1\quad g_{P_y}=i\tau ^1\quad g_I=i\tau ^3;\\
g_T&=&i\tau ^3\quad g_{P_x}=i\tau ^1\quad g_{P_y}=i\tau ^1\quad g_I=i\tau ^2;\\
g_T&=&i\tau ^3\quad g_{P_x}=i\tau ^1\quad g_{P_y}=i\tau ^2\quad g_I=i\tau ^1;\\
g_T&=&i\tau ^3\quad g_{P_x}=i\tau ^2\quad g_{P_y}=i\tau ^1\quad g_I=i\tau ^1;
\eeq
\beq
g_T&=&\tau ^0\quad g_{P_x}=i\tau ^1\quad g_{P_y}=i\tau ^2\quad g_I=i\tau ^3;
\eeq
\beq
g_T&=&i\tau ^3\quad g_{P_x}=i\tau ^1\quad g_{P_y}=i\tau ^2\quad g_I=i\tau ^3;\\
g_T&=&i\tau ^3\quad g_{P_x}=i\tau ^3\quad g_{P_y}=i\tau ^1\quad g_I=i\tau ^2;\\
g_T&=&i\tau ^3\quad g_{P_x}=i\tau ^2\quad g_{P_y}=i\tau ^3\quad g_I=i\tau ^1.
\eeq

There are 48 different gauge inequivalent choices of $g$s. Therefore, the total number of PSGs is $48\times 2^5\times 4=6144$. However, when $g_T=\tau ^0$, to acquire non-vanishing ansatze, $\eta _T$ must be identical to\cite{Wen_2002} $-1$. Therefore, $15\times 2^4\times 4=960$ PSGs are killed and the totally number of PSGs reduces to 5184.

\subsection{\label{sec:level1}Ansatzes of the Nearest-Neighbour Spin Coupling Model}

In this section we assume that only $u_{i,i+x}$, $u_{i,i+y}$, $u_{i,i+z}$, $u_{i,i-x+z}$, $u_{i,i-y+z}$ and $u_{i,i-x-y+z}$ are non-vanishing. First, an ansatz $u_{i,i+m}$ under $T_xG_x$, $T_yG_y$ and $T_zG_z$ reads
\beq
u_{i,i+m}=\eta _x^{i_xm_z}\eta _y^{i_ym_z}u_m^l\tau ^l,\quad l=0,1,2,3,
\eeq
where $u_m^i$, $i=1,2,3$ is real and $u_m^0$ is pure imaginary. $\mathcal{T}$ and $I$ further give constraints
\beq
\eta _t^{m_z}g_Tu_m^l\tau ^lg_T^{-1}&=&-u_m^l\tau ^l;\\
\eta _I^{m_z}g_Iu_m^l\tau ^lg_I^{-1}&=&u_{I(m)}^l\tau ^l.
\eeq
Using $u_{I(m)}=u_{-m}=u_m^{\dagger}$, we can conclude that when
\beq
\eta _t&=&1,\eta _I=1,g_T=i\tau ^3,\text{ and }g_I=i\tau ^3,
\eeq
all $u_m$ vanish. This kills $10\times 2^3\times 4=320$ PSGs and the totally number of PSGs is $5184-320=4864$. When
\beq
\eta _t\eta _I=-1,g_T=i\tau ^3,g_I=i\tau ^{1,2}\text{ or }\tau ^0,
\eeq
all $u_m^l$ vanish for odd $m_z$, namely only $u_{i,i+x}$ and $u_{i,i+y}$ remain non-zero. These ansatzes degenerate to describe spin liquids in a rectangular lattice in 2D plane, which is irrelevant to us. There is another similar case. When $G_x=(-)^{i_z}\tau ^0$, $G_{P_x}$ will give the constraint
\beq
(\cos(\frac{\pi}{2}m_z)+(-)^{i_z}\sin(\frac{\pi}{2}m_z))\cdot \nonumber\\
\eta _p^{m_y}g_{P_x}u_m^l\tau ^lg_{P_x}^{-1}=u_{P_x(m)}^l\tau ^l.
\eeq
For odd $m_z$, $l.h.s.$ is a function of $i_z$ while $r.h.s.$ is not, which indicates that all the $u_m^l$ for odd $m_z$ must vanish to satisfy the equation. As indicated in the previous argument, we do not take consideration of these ansatzes. Therefore, only $G_x(i)=G_y(i)=\tau ^0$ case will be under consideration. The number of PSGs left is $(4864-(9+14)\times 2^3\times 4)\div 4=1032$.

When $G_x(i)=G_y(i)=\tau ^0$, $P_x$ and $P_y$ give constraints on ansatzes
\beq
g_{P_x}u_x^l\tau ^lg_{P_x}^{-1}&=&u_x^{l\dagger}\tau ^l;\\
\eta _pg_{P_x}u_y^l\tau ^lg_{P_x}^{-1}&=&u_y^l\tau ^l;\\
\eta _{5px}g_{P_x}u_z^l\tau ^lg_{P_x}^{-1}&=&u_{-x+z}^l\tau ^l;\\
\eta _p\eta _{5px}g_{P_x}u_{-y+z}^l\tau ^lg_{P_x}^{-1}&=&u_{-x-y+z}^l\tau ^l;
\eeq
\beq
\eta _pg_{P_y}u_x^l\tau ^lg_{P_y}^{-1}&=&u_x^l\tau ^l;\\
g_{P_y}u_y^l\tau ^lg_{P_y}^{-1}&=&u_y^{l\dagger}\tau ^l;\\
\eta _{5py}g_{P_y}u_z^l\tau ^lg_{P_y}^{-1}&=&u_{-y+z}^l\tau ^l;\\
\eta _p\eta _{5py}g_{P_y}u_{-x+z}^l\tau ^lg_{P_y}^{-1}&=&u_{-x-y+z}^l\tau ^l.
\eeq
These equations determine the constrains of ansatzes in numerical calculation. Two of the constrains are employed. One is the periodic condition, that the periodicity of all the ansatzes is 1, and the other is the sector condition, that the ansatzes satisfy
\beq
u_z=s_x u_{-x+z}=s_y u_{-y+z}=s_{xy} u_{-x-y+z},
\eeq
with $s_x,s_y,s_{xy}=\pm1$. According to numerical results, there are at most 311 inequivalent ansatzes.

\section{Numerical Method and Data for Strong Coupling Case}

Differential evolution (DE), originally developed by Storn and Price\cite{Storn_1997}, is a meta-heuristic algorithm that globally optimizes a given objective function in an iterative manner. Usually the objective function is treated as a black box and no assumptions are needed. For example, unlike traditional gradient decent, conjugate gradient and qusai-Newton methods, derivatives are not needed. Evaluation of derivatives of mean-field energy defined previously is time-consuming for which DE is suitable. Besides, another algorithm, the Nelder-Mead method is also tested but doesn't perform as good as DE.

DE works with a group (called population) of solution candidates (called agents), which is initialized randomly. In each iterative step, a certain agent is selected and a new agent is generated from this agent and two other randomly selected agents in a random, linear way. If the new agent is better that the old agent, the old agent is replaced by the new one. If not, the trial agent is simply discarded. This procedure continues until some certain accuracy is reached. 

In this paper, DE is used to optimize the mean-field energy with respect to ansatzes. Constrained by PSG's, number of optimizing variables is restricted to be 12. The number of agents is set to be 120, 10 times the number of variables, with differential weight being 0.9 and cross over probability being 0.5. 

\subsection{Fourier Transformation of the Mean Field Hamiltonian}

The mean field Hamiltonian reads
\beq
H_{MF}=H_{MF}^f+H_{MF}^b
\eeq
with 
\beq
H_{MF}^f=\frac{3}{8} \sum_{\vec{r}} \sum_\alpha[
\psi_{\alpha_1}^\dagger(\vec{r})U_\alpha \psi_{\alpha_2}(\vec{r}) \nonumber \\
+\psi_{\alpha_2}^\dagger(\vec{r})U_\alpha^\dagger \psi_{\alpha_1}(\vec{r})
]
\eeq
and
\beq
H_{MF}^b=\frac{t}{2} \sum_{\vec{r}}  \sum_\alpha[
b_{\alpha_1}^\dagger(\vec{r})U_\alpha b_{\alpha_2}(\vec{r})\nonumber \\
+b_{\alpha_2}^\dagger(\vec{r})U_\alpha^\dagger b_{\alpha_1}(\vec{r})] 
\eeq

Here superscripts $f$ and $b$ mean fermion and boson respectively. $\vec{r}$ refers to the coordinate of one certain super-cell. $\alpha$ refers to the index of one certain bond in a cell. $\alpha_1$ is index of the first end of bond $\alpha$, $\alpha_2$ is the other end. $U_\alpha$ is a $2 \times 2$ matrix containing various $u_{ij}$ so that above formulas are consistent with equation \ref{eq:51}. In this holon-condensed case, bosons are treated as scalars.

Only derivation of Fourier-transformed form for the fermion Hamiltonian is shown in details. The Fourier-transformed form of the boson Hamiltonian can be obtained by just replacing $\psi$ with $b$ since commutation relations are not included in derivation.

Take substitutions
\beq
\psi_{\alpha_1}(\vec{r})=\frac{1}{\sqrt{N}}\sum_{\vec{k}}{\psi_{\alpha_1}(\vec{k})}e^{\mathbbm{i}\vec{k}\cdot (\vec{r}+\vec{l}_{\alpha_1})}
\eeq
and
\beq
\psi_{\alpha_2}(\vec{r})=\frac{1}{\sqrt{N}}\sum_{\vec{k}}{\psi_{\alpha_2}(\vec{k})}e^{\mathbbm{i}\vec{k}\cdot (\vec{r}+\vec{l}_{\alpha_2})},
\eeq
we further have 
\beq
H_{MF}^f=\frac{3}{8}\sum_{\vec{k}}\sum_{\alpha}[
\psi_{\alpha_1}^\dagger(\vec{k})U_\alpha e^{\mathbbm{i}\vec{k}\cdot (\vec{l}_{\alpha_2}-\vec{l}_{\alpha_1})} \psi_{\alpha_2}&(\vec{k}) \nonumber \\
+\psi_{\alpha_2}^\dagger(\vec{k})U_\alpha^\dagger e^{-\mathbbm{i}\vec{k}\cdot (\vec{l}_{\alpha_2}-\vec{l}_{\alpha_1})} \psi_{\alpha_1}(\vec{k})
].&
\eeq

It should be noted that this Hamiltonian is block-diagonalized with respect to $\vec{k}$. So we can calculate eigenvalues and eigenvectors of each block-matrix individually to reduce calculation workload.

\beq
\mathcal{H}_{MF}^b=\frac{t}{2}\sum_{\vec{k}}\sum_{\alpha}[
b_{\alpha_1}^\dagger(\vec{k})U_\alpha e^{\mathbbm{i}\vec{k}\cdot (\vec{l}_{\alpha_2}-\vec{l}_{\alpha_1})} b_{\alpha_2}&(\vec{k}) \nonumber \\
+b_{\alpha_2}^\dagger(\vec{k})U_\alpha^\dagger e^{-\mathbbm{i}\vec{k}\cdot (\vec{l}_{\alpha_2}-\vec{l}_{\alpha_1})} b_{\alpha_1}(\vec{k})
].&
\eeq

These two equations can be rephrased in matrix form:
\beq
H_{MF}^f=\sum_{\vec{k}}{\psi^\dagger(\vec{k}) Q_{MF}^f(\vec{k}) \psi(\vec{k})}
\eeq
with
\beq
\psi({\vec{k}})=
\begin{pmatrix}
\psi_{1,\uparrow}(\vec{k})\\
\psi_{2,\uparrow}(\vec{k})\\
\vdots \\
\psi_{N_{site},\uparrow}(\vec{k})\\
\psi_{1,\downarrow}(\vec{k}) \\
\vdots \\
\psi_{N_{site},\downarrow}(\vec{k})
\end{pmatrix}.
\eeq

Here $N_{site}$ means the number of sites in one unit cell.

The $Q_{MF}^f$ can be diagonalized as 
\beq
Q_{MF}^f(\vec{k})=S^f(\vec{k})D^f(\vec{k})S^{f \dagger}(\vec{k}).
\eeq
Denote $\phi(\vec{k})=S^{f \dagger}(\vec{k}) \psi(\vec{k})$, we further have 
\beq
H_{MF}^f=\sum_{\vec{k}}\sum_{i=1}^{2N_{site}}{\lambda_i^f(\vec{k})\phi_i^\dagger(\vec{k})\phi_i(\vec{k})}.
\eeq
To obtain energy of the original Hamiltonian, we need to evaluate the average $\langle \psi_i^\dagger(\vec{k_1})\psi_j(\vec{k_2}) \rangle_0$. The subscript $0$ means that average is taken in a Gaussian level. These two averages can be expressed as
\begin{equation}
\langle \psi_i^\dagger(\vec{k_1})\psi_j(\vec{k_2}) \rangle_0=
\delta_{\vec{k_1},\vec{k_2}} \sum_{l=1}^{2N_{site}}{S_{il}^{f*}S_{jl}^f \langle \phi_l^\dagger(\vec{k_1})\phi_l(\vec{k_1}) \rangle_0 },
\end{equation}
where
\begin{equation}
\langle \phi_l^\dagger(\vec{k})\phi_l(\vec{k}) \rangle_0=
\begin{cases}
0, &\lambda_l^f(\vec{k})>0 \\
1, &\lambda_l^f(\vec{k})<0.
\end{cases}
\end{equation}

For simplicity, we define several functions:
\beq
n_f(i,s)=\sum_{\vec{k}}\langle \psi_{i,s}^\dagger(\vec{k})  \psi_{i,s}(\vec{k}) \rangle_0,
\eeq
\beq
n_b(i,s)=\sum_{\vec{k}} b_{i,s}^\dagger(\vec{k})  b_{i,s}(\vec{k}) ,
\eeq

\begin{equation}
O_f(\alpha,s_1,s_2)=\sum_{\vec{k}}e^{\mathbbm{i}\vec{k} \cdot (\vec{l}_{\alpha_2}-\vec{l}_{\alpha_1})} \langle \psi_{\alpha_1,s_1}^\dagger(\vec{k}) \psi_{\alpha_2,s_2}(\vec{k}) \rangle_0,
\end{equation}

\begin{equation}
O_b(\alpha,s_1,s_2)=\sum_{\vec{k}}e^{\mathbbm{i}\vec{k} \cdot (\vec{l}_{\alpha_2}-\vec{l}_{\alpha_1})}  b_{\alpha_1,s_1}^\dagger(\vec{k}) b_{\alpha_2,s_2}(\vec{k}) .
\end{equation}

Here $i$ is the site index in one super-cell, $s$, $s_1$ and $s_2$ can be $\uparrow$ or $\downarrow$.

\subsection{Evaluation of the Energy}

With assistance with definitions above, energy of the original Hamiltonian can be expressed as
\beq
\langle H \rangle_0 =\langle H_1 \rangle_0-t\langle H_2 \rangle_0,
\eeq

\beq
\langle H_1 \rangle_0=\sum_{\alpha}[
\frac{N_{cell}}{4}-\frac{1}{4}n_f(\alpha_1,\uparrow)-\frac{1}{4}n_f(\alpha_1,\downarrow) \nonumber \\
-\frac{1}{4}n_f(\alpha_2,\downarrow)-\frac{1}{4}n_f(\alpha_2,\downarrow) \nonumber \\
-\frac{1}{2N_{cell}}O_f(\alpha,\downarrow,\uparrow)O_f(\alpha,\uparrow,\downarrow) \nonumber \\
+\frac{1}{2N_{cell}}O_f(\alpha,\uparrow,\uparrow)O_f(\alpha,\downarrow,\downarrow) \nonumber \\
-\frac{1}{2N_{cell}}O_f^*(\alpha,\downarrow,\uparrow)O_f^*(\alpha,\uparrow,\downarrow) \nonumber \\
+\frac{1}{2N_{cell}}O_f^*(\alpha,\uparrow,\uparrow)O_f^*(\alpha,\downarrow,\downarrow) \nonumber \\
-\frac{1}{4N_{cell}}O_f^*(\alpha,\uparrow,\uparrow)O_f(\alpha,\uparrow,\uparrow) \nonumber \\
+\frac{1}{4N_{cell}}n_f(\alpha_1,\uparrow)n_f(\alpha_2,\uparrow) \nonumber \\
-\frac{1}{4N_{cell}}O_f^*(\alpha,\uparrow,\downarrow)O_f(\alpha,\uparrow,\downarrow) \nonumber \\
+\frac{1}{4N_{cell}}n_f(\alpha_1,\uparrow)n_f(\alpha_2,\downarrow) \nonumber \\
-\frac{1}{4N_{cell}}O_f^*(\alpha,\downarrow,\uparrow)O_f(\alpha,\downarrow,\uparrow) \nonumber \\
+\frac{1}{4N_{cell}}n_f(\alpha_1,\downarrow)n_f(\alpha_2,\uparrow) \nonumber \\
-\frac{1}{4N_{cell}}O_f^*(\alpha,\downarrow,\downarrow)O_f(\alpha,\downarrow,\downarrow) \nonumber \\
+\frac{1}{4N_{cell}}n_f(\alpha_1,\downarrow)n_f(\alpha_2,\downarrow)
],
\eeq
and
\beq
\langle H_2 \rangle_0 =\frac{1}{2N_{cell}}\sum_{\alpha}[
O_b(\alpha,\uparrow,\downarrow)O_f(\alpha,\downarrow,\uparrow) \nonumber \\
+O_b(\alpha,\downarrow,\uparrow)O_f(\alpha,\uparrow,\downarrow) \nonumber \\
-O_b(\alpha,\uparrow,\uparrow)O_f(\alpha,\downarrow,\downarrow) \nonumber \\
-O_b(\alpha,\downarrow,\downarrow)O_f(\alpha,\uparrow,\uparrow) \nonumber \\
+O_b^*(\alpha,\uparrow,\downarrow)O_f^*(\alpha,\downarrow,\uparrow) \nonumber \\
+O_b^*(\alpha,\downarrow,\uparrow)O_f^*(\alpha,\uparrow,\downarrow) \nonumber \\
-O_b^*(\alpha,\uparrow,\uparrow)O_f^*(\alpha,\downarrow,\downarrow) \nonumber \\
-O_b^*(\alpha,\downarrow,\downarrow)O_f^*(\alpha,\uparrow,\uparrow) \nonumber \\
+O_b(\alpha,\uparrow,\uparrow)O_f^*(\alpha,\uparrow,\uparrow) \nonumber \\
+O_b(\alpha,\uparrow,\downarrow)O_f^*(\alpha,\uparrow,\downarrow) \nonumber \\
+O_b(\alpha,\downarrow,\uparrow)O_f^*(\alpha,\downarrow,\uparrow) \nonumber \\
+O_b(\alpha,\downarrow,\downarrow)O_f^*(\alpha,\downarrow,\downarrow) \nonumber \\
+O_b^*(\alpha,\uparrow,\uparrow)O_f(\alpha,\uparrow,\uparrow) \nonumber \\
+O_b^*(\alpha,\uparrow,\downarrow)O_f(\alpha,\uparrow,\downarrow) \nonumber \\
+O_b^*(\alpha,\downarrow,\uparrow)O_f(\alpha,\downarrow,\uparrow) \nonumber \\
+O_b^*(\alpha,\downarrow,\downarrow)O_f(\alpha,\downarrow,\downarrow)
].
\eeq

\section{Construction of the Interaction Hamiltonian in Weak Coupling Limit}

A generic interaction term reads (spin indices neglected)\cite{Lin}
\beq
H'=\int\prod_a\frac{k_a}{2\pi}\sum_{P_a,i_a}\delta(Q)V({P_a,i_a,k_a})\nonumber\\
\psi_{i_1}^{P_1\dag}(k_1)\psi_{i_2}^{P_2\dag}(k_2)\psi_{i_3}^{P_3}(k_3)\psi_{i_4}^{P_4}(k_4),
\eeq
where $P=1(-1)$ for R(L). The constraint of momentum conservation
\beq
0=Q=&-&P_1k_{Fi_1}-P_2k_{Fi_2}+P_3k_{Fi_3}+P_4k_{Fi_4}\nonumber\\
&-&k_1-k_2+k_3+k_4.
\eeq
As the momentum of chiral fermions ($k_i$) are much smaller than Fermi vectors ($k_{Fi}$), the momentum conservation is reduced to
\beq
-P_1k_{Fi_1}-P_2k_{Fi_2}+P_3k_{Fi_3}+P_4k_{Fi_4}=0.
\eeq
Therefore, only two types of interactions are allowed. The first one is intra-band scattering $\psi_{i}^{R\dag}\psi_{j}^{L\dag}\psi_{j}^{L}\psi_{i}^{R}$ and the second one is inter-band scattering $\psi_{i}^{R\dag}\psi_{i}^{L\dag}\psi_{3-i}^{L}\psi_{3-i}^{R}$. The purely chiral terms like $\psi_{i}^{R\dag}\psi_{j}^{R\dag}\psi_{j}^{R}\psi_{i}^{R}$ do not generate renormalization at leading order\cite{Lin} and thus are neglected. When spin is included, we define charge and spin current
\beq
T_{ij}&=&\sum_{s}\psi_{i,s}^{\dag}\psi_{i,s},\\
\vec{T}_{ij}&=&\frac{1}{2}\sum_{s,s'}\psi_{i,s}^{\dag}\vec{\sigma}_{ss'}\psi_{i,s'},
\eeq
and the interaction Hamiltonian density in real space can be written as
\beq
\mathcal{H'}&=&f_{ij}^{\rho}T_{ii}^R T_{jj}^L-f_{ij}^{\sigma}\vec{T}_{ii}^R\cdot\vec{T}_{jj}^L\nonumber\\
&+&{f'}_{ii}^{\rho}T_{i,3-i}^RT_{i,3-i}^L-{f'}_{i,i}^{\sigma}\vec{T}_{i,3-i}^R\cdot \vec{T}_{ij}^L,
\eeq

\section{Derivation of RG Equations}

Define $z_i=v_i\tau -ix$. Determined by the operator product expansion (OPE) in terms of chiral fermions
\beq
\psi_{i,s}^R(x,\tau)\psi_{j,s'}^{R\dag}(0,0)\sim\frac{\delta_{ij}\delta{ss'}}{2\pi z_i},\\
\psi_{i,s}^L(x,\tau)\psi_{j,s'}^{L\dag}(0,0)\sim\frac{\delta_{ij}\delta{ss'}}{2\pi z_i^*},
\eeq
the current algebra reads\cite{Lin}
\beq
T_{ij}^R(x,\tau)T_{lm}^R(0,0)&\sim& \frac{\delta_{il}}{2\pi z_j}T_{jm}^R-\frac{\delta_{jm}}{2\pi z_i}T_{il}^R,\\
T_{ij}^{Ra}(x,\tau)T_{lm}^{Rb}(0,0)&\sim& \frac{\delta_{ab}}{4}(\frac{\delta_{il}}{2\pi z_j}T_{jm}^R-\frac{\delta_{jm}}{2\pi z_i}T_{il}^R)\\
&+&\frac{i\epsilon_{abc}}{2}(\frac{\delta_{il}}{2\pi z_j}T_{jm}^{Rc}+\frac{\delta_{jm}}{2\pi z_i}T_{il}^{Rc}),\nonumber\\
T_{ij}^{Ra}(x,\tau)T_{lm}^R(0,0) &\sim& \frac{\delta_{il}}{2\pi z_j}T_{jm}^{Ra}-\frac{\delta_{jm}}{2\pi z_i}T_{il}^{Ra},
\eeq
where $T^{Ra}$ is the components of the vector current $\vec{T}^R$. For $T^L$, the current algebra is the same, except $z_i\rightarrow z_i^*$. Employing the standard method\cite{Fradkin} and using the current algebra above, we obtain the RG equations
\beq
\dot{f}_{ii}^{\rho}&=&-\frac{16({f'}_{ii}^{\rho})^2+3({f'}_{ii}^{\sigma})^2}{32\pi v_i},\\
\dot{f}_{ii}^{\sigma}&=&-\frac{2(f_{ii}^{\sigma})^2+4{f'}_{ii}^{\rho}{f'}_{ii}^{\sigma}+({f'}_{ii}^{\sigma})^2}{4\pi v_i},\\
\dot{f}_{i,3-i}^{\rho}&=&\frac{16({f'}_{ii}^{\rho})^2+3({f'}_{ii}^{\sigma})^2}{16\pi(v_i+v_{3-i})},\\
\dot{f}_{i,3-i}^{\sigma}&=&\frac{2(f_{i,3-i}^{\sigma})^2+4{f'}_{ii}^{\rho}{f'}_{ii}^{\sigma}-({f'}_{ii}^{\sigma})^2}{2\pi(v_i+v_{3-i})},\\
\dot{f'}_{ii}^{\rho}&=&\frac{16f_{ij}^{\rho}{f'}_{ii}^{\rho}+3f_{ij}^{\sigma}{f'}_{ii}^{\sigma}}{8\pi(v_i+v_{3-i})}\nonumber\\
&-&\sum_i \frac{16f_{ii}^{\rho}{f'}_{ii}^{\rho}+3f_{ii}^{\sigma}{f'}_{ii}^{\sigma}}{32\pi v_i},\\
\dot{f'}_{ii}^{\sigma}&=&\frac{2f_{ij}^{\rho}{f'}_{ii}^{\sigma}+2f_{ij}^{\sigma}{f'}_{ii}^{\rho}-f_{ij}^{\sigma}{f'}_{ii}^{\sigma}}{\pi(v_i+v_{3-i})}\nonumber\\
&-&\sum_i\frac{2f_{ii}^{\rho}{f'}_{ii}^{\sigma}+2f_{ii}^{\sigma}{f'}_{ii}^{\rho}+f_{ii}^{\sigma}{f'}_{ii}^{\sigma}}{4\pi v_i}.
\eeq
Symmetries of the coupling constants are employed in the derivation of the equations above. The initial value of these coupling constants are
\beq
f_{ij,0}^{\rho}&=&\frac{1}{4} f_{ij}^{\sigma}=J'(1-(-1)^{i+j}\cos(\frac{k_{Fi}+k_{Fj}}{2}))\nonumber\\
&+&\frac{1}{2}J(1-\cos(k_{Fi}+k_{Fj}))+\delta_{ij}J_y,\\
{f'}_{ii,0}^{\rho}&=&\frac{1}{4} {f'}_{ii}^{\sigma}=J\sin(k_{F1})\sin(k_{F2})\nonumber\\
&-&2J'\sin(\frac{k_{F1}}{2})\sin(\frac{k_{F2}}{2})-J_y.
\eeq
The RG flows are calculated numerically.

\section{Bosonization Dictionary\cite{Fradkin}}

The bosonization dictionary\cite{Fradkin} reads
\beq
\psi_{i,s}^{R/L}\sim\eta_{i,s}e^{i\sqrt{4\pi}\phi^{R/L}_{i,s}},
\eeq
where the chiral boson fields satisfy commutation relation\cite{Lin}
\beq
[\phi^{R}_{i,s}(x),\phi^{R}_{i',s'}(y)]&=&-[\phi^{L}_{i,s}(x),\phi^{L}_{i',s'}(y)]\nonumber\\
&=&\frac{i}{4}\text{sgn}(x-y)\delta_{ii'}\delta_{ss'},
\eeq
\beq
[\phi^{R}_{i,s}(x),\phi^{L}_{i',s'}(y)]&=&\frac{i}{4}\delta_{ii'}\delta_{ss'},
\eeq
and $\eta$s are Klein factors satisfying $\{\eta_{i,s},\eta_{i',s'}\}=2\delta_{ii'}\delta_{ss'}$. To describe spin and charge modes, we further define
\beq
\phi_{i,\rho}=\frac{1}{\sqrt{2}}(\phi^R_{i,\uparrow}+\phi^R_{i,\downarrow}+\phi^L_{i,\uparrow}+\phi^L_{i,\downarrow}),\\
\theta_{i,\rho}=\frac{1}{\sqrt{2}}(\phi^R_{i,\uparrow}+\phi^R_{i,\downarrow}-\phi^L_{i,\uparrow}-\phi^L_{i,\downarrow}),\\
\phi_{i,\sigma}=\frac{1}{\sqrt{2}}(\phi^R_{i,\uparrow}-\phi^R_{i,\downarrow}+\phi^L_{i,\uparrow}-\phi^L_{i,\downarrow}),\\
\theta_{i,\sigma}=\frac{1}{\sqrt{2}}(\phi^R_{i,\uparrow}-\phi^R_{i,\downarrow}-\phi^L_{i,\uparrow}+\phi^L_{i,\downarrow}),
\eeq
where subscript $\rho$ represents charge mode and $\sigma$ represents spin mode, respectively.

\bibliographystyle{apsrev4-1}

\bibliography{MC.bib}
\end{document}